\documentstyle[prd,aps,preprint]{revtex}

\begin{document}

\draft
\tighten
\widetext

\preprint{
          McGill/95--14     \hspace*{10mm}
          HEPHY--PUB 620/95 \hspace*{10mm}
          MSUCL--974        \hspace*{10mm}
          hep-ph/9510409}

\title{Dilepton bremsstrahlung from pion-pion scattering
in a relativistic OBE model}

\author{H.\ C.\ Eggers}

\address{Department of Physics, McGill University,
Montr\'eal, Qu\'ebec H3A 2T8, Canada \\
and
Institut f\"ur Hochenergiephysik der \"Osterreichischen
Akademie der Wissenschaften, \\
Nikolsdorfergasse 18, A--1050 Vienna, Austria}

\author{R.\ Tabti, C.\ Gale}

\address{Department of Physics, McGill University,
Montr\'eal, Qu\'ebec H3A 2T8, Canada }

\author{K.\ Haglin}

\address{
National Superconducting Cyclotron Laboratory, Michigan State University\\
East Lansing, MI 48824-1321, USA}

\date{October 1995}

\maketitle

\begin{abstract}
We have made a detailed and quantitative study of dilepton production
via bremsstrahlung of a virtual photon during pion-pion collisions.
Most calculations of electromagnetic radiation from strong interaction
processes rely on the soft photon approximation (SPA).  The  conditions
underlying this approximation are generally violated when dilepton
spectra are calculated in terms of their invariant mass, so that an
approach going beyond the SPA becomes necessary.  Superseding
previous derivations, we derive an exact formula for the bremsstrahlung
cross section.  The resulting formulation is compared to various forms
based on the SPA, the two-particle phase space approximation and
R\"uckl's formula using a relativistic One Boson Exchange
(OBE) model.  Within the OBE approach, we show that approximations to the
bremsstrahlung dilepton cross sections often differ greatly from the
exact result; discrepancies become greater both with rising temperature
and with invariant mass.  Integrated dilepton production rates are
overestimated by R\"uckl-based approximations by factors 1.5--8.0.  The
largest discrepancies occur for the reaction $\pi^+\pi^+ \to
\pi^+\pi^+\ell^+\ell^-$, where such approximations overestimate the
exact rate by factors ranging from 2 to 30 for invariant masses between
10 and 500 MeV.  Our findings, combined with recent estimates of the
Landau-Pomeranchuk effect, indicate that bremsstrahlung dileptons rates
in ultrarelativistic heavy ion collisions should be even more
suppressed than had been thought before.
\end{abstract}

\pacs{13.40.-f, 25.75.-q, 13.75.Lb, 25.80.Ek}

\hyphenation{brems-strah-lung}

\section{Introduction}
\label{sec:intro}

Dileptons produced in heavy ion collisions interact only
electromagnetically with their hadronic surroundings.  Especially the
early phases of such collisions, when dilepton production rates are
largest, can therefore be explored with lepton pairs.  Some of the
initial ideas on photon \cite{Fein76} and dilepton \cite{Shu78a}
\cite{Kaj86a} emission have met with considerable interest, and this
area has now expanded into a major component of research in
ultrarelativistic heavy ion collisions \cite{qm95}.  On the
experimental side, dilepton experiments for hadronic collisions
\cite{Ada83a}--\cite{Roc93a} have been
complemented only recently by nucleus-nucleus experiments at very high
energies \cite{HELIOS94a}--\cite{Tse95a}.
Since the spectrum of lepton pairs and of real photons
has been proposed as a signature of the quark-gluon plasma \cite{qm95},
it is imperative quantitatively to understand the emission mechanisms
from the confined sector of QCD, which in this sense constitutes the
background.

Guided by theoretical estimates, we can roughly divide the lepton pair
production cross section, expressed in terms of its invariant mass
$M$,  into a number of distinct regions. For $M >  m_{J/\psi}$, the
Drell-Yan contribution dominates, while pairs arising from the
semileptonic decay of $D \bar{D}$ are also important \cite{Vogt94}.
For lepton pair invariant masses around the light vector meson masses
($\rho$, $\omega$, and $\phi$), two-body \cite{Gal94} and three-body
\cite{Lic94a} reactions dominate. The latter processes also extend
their influence to the region $m_\phi  < M <  m_{J/\psi}$, where  the
quark-gluon plasma may become visible for sufficiently high initial
temperatures \cite{Wong94}. In this paper, we concentrate on the
regions of low invariant masses, the so-called ``soft'' limit $M \leq
300$ MeV. There, production via bremsstrahlung is expected to play an
important role and many authors have recently concerned themselves with
such reactions
\cite{Gal87a}--\cite{Kno95a}.

Since particle production plays an important role in ultrarelativistic
heavy ion collisions, we consider only events that arise from
microscopic meson-meson dynamics. In high energy heavy ion experiments,
the meson to baryon ratio is such that the absence of baryons in our
treatment is not expected to constitute a major hurdle to
phenomenology. More specifically, we restrict our studies to those of
pion-pion bremsstrahlung.

The production of photons or dileptons necessarily involves 3- or
4-particle final states. Except for special cases such as cross
sections at fixed $\bbox{q}$ \cite{Hag91a}, such final states have
universally been handled under some approximation. It is our goal in
this paper to go beyond these by presenting an essentially exact
formulation for the $\pi\pi \to \pi\pi\ell^+ \ell^-$ cross section (as
a function of dilepton invariant mass) and, in the process, to test how
far these approximations may be trusted. In order to achieve these
goals, we formulate the pion-pion interaction in terms of a
relativistic One Boson Exchange model (OBE).

Note that we do not attempt to compare directly to experimental data
because the main point at issue is a comparison of theoretical
approaches to bremsstrahlung.
To do justice to small-$M$ data would require spacetime integration
over the fireball region as well as inclusion of Dalitz decays and
the Landau-Pomeranchuk effect, a substantial task which will be
postponed to future work.

Our paper is organized as follows: in the next section, we discuss the
set of approximations known under the collective banner of the Soft
Photon  Aproximation (SPA). Next, we derive exact formulae for the
bremsstrahlung generation of lepton pairs. Section \ref{sec:threesp} is
devoted to technical issues having to do with phase space; Sections
\ref{sec:curint} and \ref{sec:xsexact} then derive cross sections based
respectively on various SPA variants and the exact formulation.
Our OBE model is presented in Section \ref{sec:fullobe}; finally,
Section \ref{sec:results} contains our numerical results in the form
of cross sections $d\sigma(s)/dM$ and
rates $dN/d^4x\, dM$ and a discussion on the issues raised.

\section{The soft photon approximation}
\label{sec:low}

The Soft Photon Approximation (SPA) has been widely used in calculating
bremsstrahlung dilepton spectra. It is based on the early realization
\cite{bd64} that the cross section for production of low-energy real
photons is dominated by the corresponding hadronic amplitude. In practice,
this means neglecting the photon momentum in the hadronic matrix
element as well as all photon emission from vertices and internal lines
(see Section \ref{sec:spa}).

\subsection{Limitations}
\label{sec:limits}

In order for the SPA to be a valid approximation,
two conditions \cite{Low58a} must be met:
\begin{enumerate}
\item
The photon energy must be much smaller than the energy of any
one of the hadrons participating in the scattering,
\begin{equation}
\label{iaa}
q_0 / E \ll 1\,,
\end{equation}
\item
the process of radiating the photon must be separable from the hadronic
interaction process, i.e.\ the hadronic and electromagnetic time and
distance scales must be sufficiently different to permit separate
treatment.  The range of the hadron-hadron interaction is given roughly
by $b = m_Y^{-1}$, the inverse mass of the exchange boson, while the
distance the hadron may propagate with off-shell energy $\Delta E =
q_0$ before emitting the photon is about
$\Delta x = v/\Delta E = |\bbox{p}|/E q_0$.
The SPA is valid only when $\Delta x / b \gg 1$, or
\begin{equation}
\label{iac}
q_0 \ll m_Y |\bbox{p}|/E \,.
\end{equation}
\end{enumerate}
Implicit in these equations is, of course, a specific Lorentz frame
with respect to which the energies are measured. For the special case
where two hadrons of equal mass $m$ collide, these conditions can be
re-written in their cms frame as
\begin{eqnarray}
\label{iad}
q_0^* &\ll& \sqrt{s}/2 \,, \\
\label{iae}
q_0^* &\ll& m_Y \sqrt{1 - 4 m^2/s} \,.
\end{eqnarray}
In a simple bremsstrahlung experiment, these limits are easily
satisfied by selecting only photons or dileptons of low energy in the
laboratory frame. An early dilepton paper by R\"uckl, for example
\cite{Ruc76a}, calculated spectra for $\sqrt{s} = 27$ and $53$ GeV
while restricting transverse momentum to below 500 MeV.

In the complex multiparticle systems formed in the course of
nucleus-nucleus collisions, however, the situation is more
complicated:  There are many binary collisions, and their respective
cms frames do not generally coincide either with one another or with
the overall nucleus-nucleus cms frame.  A  photon that is soft in a
particular hadron-hadron cms therefore does not have to be soft in the
laboratory and vice versa. (This fact has to be taken into account when
addressing issues related to the spectrum of real photons
\cite{alam93}.)

In a situation where there are many cms frames, it therefore becomes
necessary to look at relativistically invariant quantities.  For lepton
pairs, the extra degree of freedom provided by a non-vanishing
invariant mass is a suitable variable of choice.  Looking at invariant
masses means, however, that $q_0$ is no longer fixed but must vary over
its full kinematic range, which for our example of colliding equal-mass
hadrons is given by (see Section \ref{sec:curint})
\begin{equation}
\label{iah}
M \leq q_0 \leq {s - 4 m^2 + M^2 \over 2\sqrt{s}} \,.
\end{equation}
Once the upper kinematic limit for $q_0$ approaches or exceeds the
bounds set by either Eq.\ (\ref{iad}) or (\ref{iae}), the
Soft Photon
Approximation clearly fails.

In Figure 1, we show the three functions (\ref{iad}) and (\ref{iae})
and (\ref{iah}) for the case $m = m_\pi = 140$ MeV, $m_Y = m_\sigma
\simeq 500$ MeV and dilepton invariant masses $M = 10$ and $300$ MeV.
It is immediately clear that the assumptions underlying the
SPA are not fulfilled even for small $M$: the kinematic range
accessible to $q_0$ is never much smaller than the limits set by the
SPA. The situation becomes even worse for larger $M$.

Closely related to the above is a common misconception that the SPA
must necessarily be valid for small invariant masses.  The confusion
stems from the fact that, because
\begin{equation}
\label{iai}
M^2 = q_0^2 - \bbox{q}^2 \,,
\end{equation}
a small value for $q_0$ necessarily implies small $M$. However, the
reverse is not true: both $q_0$ and $\bbox{q}$ can be large even while
adding up to small $M$. An extreme counterexample is of course a hard
real photon with zero invariant mass but large energy. One cannot,
therefore, a priori expect SPA calculations to coincide with full
hadronic calculations even when $M$ is small.  Section
\ref{sec:results} provides examples of such deviations; see especially
the ratios between approximations and the exact calculation in
Figures 12 and 13 which do not approach unity for small $M$.
This point has been emphasized before \cite{Schae91a}.

We are therefore led to conclude that the SPA does not generally apply
to rates or cross sections measured in terms of the dilepton invariant
mass.  Correspondingly, doubt must be cast on the validity of the SPA
in calculations pertaining to such measurements.

That said, one may still ask {\it by how much} SPA-based calculations
differ from the more complicated true situation: is it a matter of a
few percent, or orders of magnitude? Full calculations such as
presented in Section \ref{sec:fullobe} below are cumbersome and
time-consuming, and it is therefore of value to understand
quantitatively to what percentage the SPA approximation can be
trusted.  It is this {\it quantitative} question that we attempt to
cast some light on by making a model calculation capable both of
``true'' answers and an SPA approximation.

Clearly, a single model calculation such as presented below cannot
answer this quantitative question exhaustively. Nevertheless, its
results provide at least an indicator of the reliability of the SPA:
If, within our simple OBE, dileptons can or cannot reasonably
be described within the SPA, it is possible or even likely that the
same or similar conclusions would emerge from other models.

\subsection{The SPA as series of approximations}
\label{sec:spa}

Rather than being just a single step, the soft photon approximation as
implemented in a practical context can consist of a number of stages.
Starting with the most fundamental, these are:
\begin{enumerate}
\item[(A) ]
The virtual photon producing the lepton pair is emitted only by
external legs of the hadronic reaction; emission from internal parts of
the hadronic reaction is neglected because the radiation from internal
lines forms a sub-leading contribution \cite{Low58a}.
\item[(B) ]
The dependence of the hadronic matrix element ${\cal M}_h$ on the
photon momentum $q$ is neglected.
\item[(C) ]
The photon momentum $q$ is neglected in the phase space delta functions.
\end{enumerate}
In addition, we shall be considering in this paper the effect of
\begin{enumerate}
\item[(D) ]
approximating the virtual photon current by the real photon
current, ($J^\mu = \sum (2p_i^\mu\pm q)/(2p_i\cdot q \pm  M^2)
\to      \sum p_i^\mu/(p_i\cdot q)$;
see Eqs.\ (\ref{brk}) and (\ref{brl})).
\end{enumerate}
(A) and (B) have formed the basis for most bremsstrahlung dilepton
calculations as they are hard to avoid.  The restriction of phase space
(C), on the other hand, can be circumvented fairly easily by inserting
later a phase space correction factor (Section \ref{sec:ruckl}) or by
avoiding (C) altogether using Eqs.\ (\ref{brv}), (\ref{fcm}) and Sections
\ref{sec:threesp} and \ref{sec:xsexact}.

The change from virtual to real photon currents in step (D) is not
directly dependent on (C) and can be implemented either jointly or
separately.  We shall be investigating both alternatives and
determining their impact on the overall dilepton rate. In Section
\ref{sec:fullcross} and Section \ref{sec:fullobe} and beyond, we shall
refrain from using even the basic assumptions (A) and (B) and calculate
exact cross sections and rates within the same lagrangian.

\section{Bremsstrahlung cross section}
\label{sec:bremsf}

We wish to find the cross section for a dilepton pair $\ell^+\ell^-$
emitted in the semi-elastic scattering of two pions $a$ and $b$ into
$1$ and $2$,
\begin{equation}
\label{brc}
\pi_a + \pi_b \rightarrow \pi_1 + \pi_2 + \ell^+ + \ell^- \,.
\end{equation}
The lepton pair forms via a virtual photon $\gamma^*$ with
four-momentum $q^\mu$ and mass $q^2 = M^2$ emitted from either the
central blob or external legs of the hadronic reaction
\begin{equation}
\label{brd}
\pi_a + \pi_b \rightarrow \pi_1 + \pi_2.
\end{equation}
The leading contribution will come from photon emission from external
pion legs. First, we follow Lichard \cite{Lic95a} in a leading-terms
derivation but place more emphasis on the role of phase space; the full
formalism supplanting the leading-terms derivation is presented in
Section \ref{sec:fullcross}.

\subsection{Bremsstrahlung from external legs}
\label{sec:external}

Four diagrams contribute to the leading-term cross section, one of
which is shown in Figure 2. In the notation of Figure 2, the total
matrix element can be written as products of purely hadronic reactions
${\cal M}_h$ with currents $J_i^\mu$ and a leptonic part $L_\mu$
\cite{Cra78a},
\begin{eqnarray}
\label{bre}
{\cal M}_{\ell^+ \ell^-} &=&
 e \left[
    J_a^\mu {\cal M}_h(p_a-q,p_b,p_1,p_2)
+   J_b^\mu {\cal M}_h(p_a,p_b-q,p_1,p_2) \,,
\right.  \nonumber \\   &&{}   \left.
+   J_1^\mu {\cal M}_h(p_a,p_b,p_1+q,p_2)
+   J_2^\mu {\cal M}_h(p_a,p_b,p_1,p_2+q)
\right]\;   L_\mu \;,
\end{eqnarray}
where
\begin{equation}
\label{brh}
L_\mu = {e \over M^2} \bar u(p_-)\gamma_\mu v(p_+),
\end{equation}
and, in slightly condensed notation, the terms of the pionic
current are
\begin{eqnarray}
\label{brf}
J_{a,b}^\mu &=&
{ -Q_{a,b}(2p_{a,b} - q)^\mu \over 2 p_{a,b}\cdot q - M^2} \,,
\\
J_{1,2}^\mu &=&
{ Q_{1,2}(2p_{1,2} + q)^\mu \over 2 p_{1,2}\cdot q + M^2} \,,
\end{eqnarray}
with $Q_i$ the charge of pion $i$ in units of the proton charge.
Note that $J_i^\mu$ contains the propagator for the off-shell pion.
Making approximation (B), one now neglects the dependence on $q$ of
the four hadronic matrix elements; for example
\begin{equation}
\label{bri}
{\cal M}_h(p_a-q,p_b,p_1,p_2) \sim
{\cal M}_h(p_a,  p_b,p_1,p_2)\,,
\end{equation}
permitting the usual factorization of ${\cal M}_{\ell^+\ell^-}$ into
a purely hadronic matrix element and a photon-plus-lepton part:
\begin{equation}
\label{brj}
{\cal M}_{\ell^+\ell^-} \simeq
e {\cal M}_h(p_a,p_b,p_1,p_2)
  J^\mu(p_a,p_b,p_1,p_2,q) L_\mu(p_+,p_-)\,,
\end{equation}
where now the virtual-photon current
\begin{equation}
\label{brk}
J^\mu = \sum_{i=1,2}    Q_i{(2p_i+q)^\mu \over 2p_i\cdot q + M^2}
      - \sum_{i=a,b}    Q_i{(2p_i-q)^\mu \over 2p_i\cdot q - M^2}
\end{equation}
contains the sum over the four diagrams.
Since approximations (A) and (B) are valid only for photons of small
$q$ and by implication for small $M$, the extra terms in numerator and
denominator are usually neglected (in what we called approximation (D))
to form the real-photon current
\begin{equation}
\label{brl}
I^\mu = \sum_{i=1,2}    Q_i{p_i^\mu \over p_i\cdot q}
      - \sum_{i=a,b}    Q_i{p_i^\mu \over p_i\cdot q}\,.
\end{equation}
Squaring the total matrix element
and summing over lepton polarizations $(s_+,s_-)$ yields,
using either current \cite{Lic95a},
\begin{eqnarray}
\label{brm}
\sum_{s_+, s_-} |{\cal M}_{\ell^+ \ell^-}|^2
&=& 4\pi\alpha |{\cal M}_h(p_a,p_b,p_1,p_2)|^2 J^\mu J^\nu L_{\mu\nu}
\nonumber \\
&=& |{\cal M}_h|^2 {32\pi^2\alpha^2\over M^2}
\left[ -J\cdot J - {(l\cdot J)^2 \over M^2} \right] \,,
\end{eqnarray}
where $l = p_+  -  p_-$ and the leptonic tensor is
\begin{equation}
\label{brn}
L_{\mu\nu} = \sum_{s_+, s_-} L_\mu L_\nu^*
= {8\pi\alpha \over M^4}
\left( q_\mu q_\nu - l_\mu l_\nu - M^2 g_{\mu\nu} \right) \,.
\end{equation}
At this point, $q$ is still defined only as the sum of
lepton momenta $(p_+ + p_-)$.

With $F = 4[(p_a\cdot p_b)^2 - m_a^2 m_b^2]^{1/2}$ the incoming
flux, the unpolarized cross section is an integral over
phase space of {\it four} outgoing particles,
\begin{eqnarray}
\label{bro}
d\sigma_{hh\ell^+\ell^-} &=&
\sum_{s_+, s_-} |{\cal M}_{\ell^+ \ell^-}|^2
{dR_4 \over (2\pi)^8 F} \,, \\
\label{brp}
dR_4 &\equiv& \delta^4(p_a+p_b-p_1-p_2-p_+-p_-)\;
d\tau_1\, d\tau_2\, d\tau_+\, d\tau_-
\end{eqnarray}
where we write
$d\tau_i = {d^3 \bbox{p_i} / 2 E_i}$ for short. This can be reduced to
a {\it three\/}-particle phase space integral by changing variables and
integrating over redundant degrees of freedom in the dilepton cms,
\begin{equation}
\label{brqq}
d\tau_+ d\tau_- = {1\over 8}\sqrt{1 - {4\mu^2 \over M^2}} \, dM^2\,
d\tau_q \,  d\Omega_+ \,,
\end{equation}
where the positron solid angle $\Omega_+$ must be kept due to the
$(l\cdot J)$ term, and $\mu$ is the lepton mass.
The dependence of $(l\cdot J)$ on $\Omega_+$ is integrated out,
\begin{equation}
\label{brs}
\int \left[ - J^2 - {(l\cdot J)^2 \over M^2} \right]
d\Omega_+
= {8\pi\over 3}\left(1 + {2\mu^2\over M^2}\right) (-J^2) \,,
\end{equation}
so that, using Eqs.\ \ref{brm}--\ref{brs}, we can express the
total cross section in terms of a leptonic factor $\kappa$
and a ``hadronic plus virtual photon'' cross section
$d\sigma_{hh\gamma^*}$,
\begin{equation}
\label{brt}
d\sigma_{hh\ell^+\ell^-} = \kappa(\mu^2,M^2)
\;d\sigma_{hh\gamma^*} \,,
\end{equation}
with
\begin{equation}
\label{bru}
\kappa \equiv M^2\, \tau(M^2) \equiv
{\alpha\over 3\pi} \left( 1 + {2\mu^2\over M^2}\right)
\sqrt{1 - {4\mu^2\over M^2}} \,,
\end{equation}
and where
\begin{equation}
\label{brv}
d\sigma_{hh\gamma^*} = 4 \pi\alpha\,
{dM^2 \over M^2}
(-J^2) |{\cal M}_h|^2 \, {dR_3 \over (2\pi)^5 F}
\end{equation}
is a function of {\it three\/}-particle phase space
\begin{equation}
\label{brw}
dR_3 = \delta^4(p_a+p_b-p_1-p_2-q)\,
d\tau_1\, d\tau_2\, d\tau_q \,.
\end{equation}
With the understanding that its derivation is valid under
approximations (A) and (B) only, Eq.\ (\ref{brt}) is exact and makes no
assumptions with respect to phase space. In particular, we note that
$q$ is still contained in the delta function constraining the phase
space in Eq.\ (\ref{brw}). Next, we shall show that Eq.\ (\ref{brt}) is
valid even when approximations (A) and (B) are not made but that
$d\sigma_{hh\gamma^*}$ takes on a form different from
Eq.\ (\ref{brv}).

\subsection{Emission from all diagrams}
\label{sec:fullcross}

When calculating dilepton cross sections for all possible diagrams,
including internal emission of photons, it is no longer possible to
factorize the matrix element into a hadronic part and a current as in
Eq.\ (\ref{bre}). Instead, we consider, for a given pion-pion
reaction, the full matrix element
\begin{equation}
\label{fcb}
{\cal M}(\pi\pi{\to}\pi\pi\ell^+ \ell^-)
= {\cal M}^\mu L_\mu = \sum_m {\cal M}_m^\mu L_\mu \,,
\end{equation}
where $m$ runs over all contributing diagrams. The sum of diagrams
is gauge invariant,
\begin{equation}
\label{fcc}
q_\mu {\cal M}^\mu = \sum_m q_\mu {\cal M}_m^\mu = 0 \,,
\end{equation}
but not necessarily the individual terms.  Unlike in the previous
sections ${\cal M}^\mu$ contains, besides the hadronic interaction, the
photon vertex and where applicable the external pion propagator, both
of which were previously part of the current.

Starting again with the cross section in four-phase space
(\ref{bro}) and transforming from $d\tau_+ d\tau_-$ to
$d\tau_q dM^2 d\Omega_+$ using Eq.\ (\ref{brqq}),
the unpolarized dilepton cross section is
\begin{equation}
\label{fcd}
d\sigma_{hh\ell^+ \ell^-} =
\sqrt{1 - {4\mu^2\over M^2}} \,
{dM^2 d\tau_1 d\tau_2 d\tau_q \delta(\cdots) \over 8 (2\pi)^8 F}
\int d\Omega_+ \sum_{s^+ s^-}
\left| {\cal M}_{\ell^+ \ell^-} \right|^2 \,.
\end{equation}
Squaring and summing over lepton polarizations, we obtain in
analogy to Eq.\ (\ref{brm})
\begin{equation}
\label{fce}
\sum_{s^+ s^-} \left|{\cal M}^\mu L_\mu \right|^2
= \sum_{m,n} {\cal M}_m^\mu {\cal M}_n^{*\nu} L_{\mu\nu} \,,
\end{equation}
where as before
$L_{\mu\nu}
= 8\pi\alpha (q_\mu q_\nu - l_\mu l_\nu - M^2 g_{\mu\nu})/M^4$.
In the dilepton cms, $l_0 = 0$ and
the second term in $L_{\mu\nu}$ integrates to
\begin{eqnarray}
\label{fcf}
\int d\Omega_+ l_\mu  {\cal M}_m^\mu l_\nu {\cal M}_n^{*\nu}
&=& 4 \int d\Omega_+
    (\bbox{p}_+{\cdot}\bbox{\cal M}_m)
    (\bbox{p}_+{\cdot}\bbox{\cal M}_n^*) \,,
\nonumber \\
&=& {4\pi\over 3} (M^2 - 4\mu^2)
\bbox{\cal M}_m \bbox{{\cdot} {\cal M}}_n^* \,.
\end{eqnarray}
Since $\bbox{q}=0$ in this frame,
${\cal M}_m^0 = q_\mu {\cal M}_m^\mu / q_0$ and hence in the dilepton cms
\begin{equation}
\label{fci}
\bbox{\cal M}_m \bbox{{\cdot}{\cal M}}_n^*
= q_0^{-2} (q_\mu {\cal M}_m^\mu) (q_\nu {\cal M}_n^{*\nu})
 - {\cal M}_m {\cdot} {\cal M}_n^*  \,,
\end{equation}
so, using gauge invariance (\ref{fcc}),
\begin{equation}
\label{fcj}
\sum_{mn} \bbox{\cal M}_m \bbox{{\cdot} {\cal M}}_n^* ({\rm cms})
= - \sum_{mn} {\cal M}_m {\cdot} {\cal M}_n^*  \,.
\end{equation}
Hence
\begin{equation}
\label{fck}
\int d\Omega_+\sum_{s^+ s^-} \left| {\cal M}^\mu L_\mu \right|^2
= {64 \pi^2 \alpha \over 3 M^2}
  \left( 1 + {2\mu^2\over M^2} \right)
  \left( - \sum_{m,n} {\cal M}_m {\cdot} {\cal M}_n^* \right) ,
\end{equation}
and the full cross section becomes
\begin{equation}
\label{fcl}
d\sigma_{hh\ell^+ \ell^-} = \kappa(M^2) d\sigma_{hh\gamma^*} \,,
\end{equation}
which is identical to Eq.\ (\ref{brt}), but with the virtual
photon cross section reading
\begin{equation}
\label{fcm}
d\sigma_{hh\gamma^*} =
{dM^2 \over M^2}
\left( - \sum_{mn} {\cal M}_m {\cdot} {\cal M}_n^* \right)
{dR_3 \over (2\pi)^5 F} \,.
\end{equation}
Comparing to eq.\ (\ref{brv}), we have the correspondence
$
4\pi\alpha(-J^\mu J_\mu)|{\cal M}_h|^2
\leftrightarrow
$
$
(-\sum_{mn} {\cal M}^{\mu}_m  {\cal M}^*_{n \mu})
$.

Equations (\ref{fcl}) and (\ref{fcm}), together with use of the 3-phase
space variables of Section \ref{sec:threesp} provide a fully covariant
formalism capable of exact treatment of both hadronic and
electromagnetic sectors of a given model.  Within the physical
limitations implicit in a given model, it makes no further assumptions
or approximations.  In particular, none of the approximations listed in
Section \ref{sec:spa} are needed.

\subsection{Dilepton rates}
\label{sec:dilrate}

In the context of nuclear collisions, one is interested more in overall
dilepton rates than in fixed-$s$ cross sections. Not attempting to
account fully for either quantum statistical or flow effects here, we
shall make the simplest assumption of a locally thermal Boltzmann gas
of pions at a (local) temperature $T$.  The rate of pair production of
invariant mass $M$ per unit of four-volume then is \cite{Kaj86a}
\begin{equation}
\label{rtk}
{dN^{\rm Boltz}_{\ell^+ \ell^-} \over d^4x\, dM^2}
= {g_{ab}\over 32 \pi^4} \int ds \, \lambda(s,m_a^2,m_b^2)
         {K_1(\sqrt{s}/T) \over (\sqrt{s}/T)}
  {d\sigma_{hh\ell^+ \ell^-} \over dM^2 } ,
\end{equation}
where
\begin{equation}
\label{fve}
\lambda(x,y,z) = (x-y-z)^2 - 4yz
\end{equation}
is the basic three-particle kinematic function \cite{Byc73a}, which for
equal masses $(m_a = m_b = m_1 = m_2)$ has the more familiar form
$\lambda(s,m^2,m^2) = s(s-4m^2)$, and $g_{ab} = (2S_a+1)(2S_b+1)$ is
the spin degeneracy factor which for pions is unity. $K_1$ is the
modified Bessel function.

\section{Three-particle phase space}
\label{sec:threesp}

To complete our preparatory work, we summarize the treatment of
three-particle phase space in terms of relativistic invariants.  The
relations of this section will be useful first in 3-phase space
integration of the currents (\ref{brk}) and (\ref{brl}) and later in
the exact treatment of bremsstrahlung dilepton production.

Three-particle phase space has been studied extensively and we here
merely outline the procedure; for details see Ref.\ \cite{Byc73a}.
As shown in Figure 3, the kinematics for a reaction
\begin{equation}
\label{fvb}
a + b \longrightarrow 1 + 2 + 3
\end{equation}
can be described in terms of the five invariants
$s   = (p_a + p_b)^2$,
$t_1 = (p_1 - p_a)^2$,
$s_1 = (p_1 + p_2)^2$,
$s_2 = (p_2 + p_3)^2$, and
$t_2 = (p_b - p_3)^2$.
For our purposes, we shall be using $p_3$ and $q$ interchangeably as
referring to the virtual photon, i.e.\ $E_3\equiv q_0$ and
$q^2 = m_3^2 \equiv M^2$.  The other momenta refer, as usual, to the
incoming and outgoing hadrons. Inverting, one gets
\begin{eqnarray}
2 p_a\cdot p_b &=& s - m_a^2 - m_b^2 \,,       \nonumber\\
2 p_a\cdot p_1 &=& m_a^2 + m_1^2 - t_1 \,,     \nonumber\\
2 p_a\cdot p_2 &=& s_1 + t_1 - t_2 - m_1^2 \,, \nonumber\\
2 p_b\cdot p_1 &=& s - s_2 + t_1 - m_a^2 \,,   \nonumber\\
2 p_b\cdot p_2 &=& s_2 + t_2 - t_1 - M^2 \,,   \nonumber\\
2 p_1\cdot p_2 &=& s_1 - m_1^2 - m_2^2 \,,     \\
2 p_a\cdot q &=& s - s_1 + t_2 - m_b^2 \,,     \nonumber\\
2 p_b\cdot q &=& m_b^2 + M^2 - t_2 \,,         \nonumber\\
2 p_1\cdot q &=& s - s_1 - s_2 + m_2^2 \,,     \nonumber\\
2 p_2\cdot q &=& s_2 - m_2^2 - M^2 \,.         \nonumber
\end{eqnarray}
The corresponding phase space integral is
\begin{equation}
\label{fvc}
dR_3(s) =  {\pi \over 4 \lambda^{1/2}(s, m_a^2, m_b^2) }
\int {dt_1\, ds_2\, ds_1\, dt_2 \over \sqrt{B} } \,,
\end{equation}
where the weighting is given in terms of the Cayley determinant
\begin{equation}
\label{fvd}
B =
\left|
\begin{array}{cccccc}
0 &  1      &  1      &  1      &  1      &  1       \\
1 &  0      &  m_2^2  &  s_2    &  t_1    &  m_1^2   \\
1 &  m_2^2  &  0      &  M^2    &  t_2    &  s_1     \\
1 &  s_2    &  M^2    &  0      &  m_b^2  &  s       \\
1 &  t_1    &  t_2    &  m_b^2  &  0      &  m_a^2   \\
1 &  m_1^2  &  s_1    &  s      &  m_a^2  &  0       \\
\end{array}
\right| \,.
\end{equation}
The Cayley determinant is
quadratic in any of its arguments; specifically, we use the form
\begin{equation}
\label{fvf}
B = \lambda(s,s_2,m_1^2) \; (t_2^+ - t_2)(t_2 - t_2^-) \,,
\end{equation}
where the kinematic limits on $t_2$ are given by
\begin{eqnarray}
\label{fvg}
t_2^{\pm} &=&
m_b^2 + M^2 - {1 \over \lambda(s,s_2,m_1^2)}
\left|
\begin{array}{ccc}
  2s                 &  s{+}s_2{-}m_1^2    &    s{-}s_1{+}M^2      \\
  s{+}s_2{-}m_1^2    &  2s_2               &    s_2{-}m_2^2{+}M^2  \\
  s{-}m_a^2{+}m_b^2  &  s_2{-}t_1{+}m_b^2  &    0                  \\
\end{array}
\right|
\nonumber\\
&&{}\pm\ \ {2 \over \lambda(s,s_2,m_1^2)}
\left[
G(s,t_1,s_2,m_a^2,m_b^2,m_1^2)
G(s_1,s_2,s,m_2^2,m_1^2,M^2)
\right]^{1/2} ,
\end{eqnarray}
written in terms of yet another determinant and the basic
four-particle kinematic function
\begin{equation}
\label{fvh}
G(x,y,z,u,v,w) \equiv {-} \left( 1 \over 2 \right)
\left|
\begin{array}{ccccc}
0 &  1  &  1   &  1   &  1  \\
1 &  0  &  v   &  x   &  z  \\
1 &  v  &  0   &  u   &  y  \\
1 &  x  &  u   &  0   &  w  \\
1 &  z  &  y   &  w   &  0  \\
\end{array}
\right| .
\end{equation}
The kinematic limits on $s_1$ are
\begin{eqnarray}
\label{fvi}
s_1^{\pm} = s + M^2
&-& {1\over 2 s_2} (s + s_2 - m_1^2)(s_2 + M^2 - m_2^2)
\nonumber\\
&\pm& {1\over 2 s_2}
    \lambda^{1/2}(s,s_2,m_1^2)\;
    \lambda^{1/2}(s_2,m_2^2,M^2) \,.
\end{eqnarray}
The limits on $s_2$ and $t_1$ are given by the Chew-Low plot:
for values of $t_1$ bounded by
\begin{eqnarray}
\label{fvj}
t_1^{\pm} = m_a^2 + m_1^2
&-& {1\over 2s} [s + m_a^2 - m_b^2][s + m_1^2 - (m_2 + M)^2]
\nonumber\\
&\pm& {1\over 2s}
    \lambda^{1/2}(s,m_a^2,m_b^2)\;
    \lambda^{1/2}(s,(m_2{+}M)^2,m_1^2) \,,
\end{eqnarray}
the limits on $s_2$ are
\begin{eqnarray}
\label{fvk}
s_2^{+} &=& s + m_1^2
 - {1\over 2m_a^2} [s + m_a^2 - m_b^2][m_a^2 + m_1^2 - t_1]
\nonumber\\
   &&\ \ \ \
 + {1\over 2m_a^2}
    \lambda^{1/2}(s,m_a^2,m_b^2)\;
    \lambda^{1/2}(t_1,m_a^2,m_1^2) \,, \\
\label{fvkm}
s_2^- &=& (m_2 + M)^2 \,.
\end{eqnarray}
Although not directly relevant for pions, we include for completeness
the case where the hadronic masses $m_a$ and $m_1$ are unequal. In this
case, $t_1$ may become larger than $t_1^+$ given above, and the
corresponding additional phase space is delimited by
\begin{equation}
\label{fvl}
t_1^+ \leq t_1 \leq (m_a - m_1)^2
\end{equation}
and for this additional domain, $s_2$ is limited by $s_2^{+}$ of
Eq.\ (\ref{fvk}) and the corresponding lower branch, given by a minus
sign before the $\lambda$-factors.  Clearly, $t_1$ resembles the
2-phase space invariant $t$, but they are distinct quantities due to
the presence of the additional momentum $q$.

Integration over pure phase space yields
\begin{equation}
\label{fvm}
{dR_3\over dt_1}(s,t_1)
=
\int_{s_2^-}^{s_2^+} ds_2\;
 {\lambda^{1/2}(s_2,m_2^2,M^2) \over s_2}
= F(s_2^+) - F(s_2^-),
\end{equation}
where
\begin{eqnarray}
F(s_2) = l_2
&-& (m_2^2 + M^2) \ln(s_2 + l_2 - m_2^2 - M^2)
\nonumber\\
&-& (m_2^2 - M^2) \ln(s_2 + l_2 + m_2^2 - M^2)
\nonumber\\
&-& (M^2 - m_2^2) \ln(s_2 + l_2 - m_2^2 + M^2) \,,
\end{eqnarray}
and we have written $l_2 =  \lambda^{1/2}(s_2,m_2^2,M^2)$ for short.
This yields identical results to the pure phase space formula
(\ref{icg}) below, as it should.  We note in passing that the ratio of
the kinematic domains of (3-space) $t_1$ to (2-space) $t$ is exactly
the phase space reduction factor found in (\ref{icj}) below.

\section{Angle-averaged-current approximations}
\label{sec:curint}

All the elements for our calculation are now in place. In this section,
we look at three approximations of increasing sophistication.  We
proceed first to find explicit expressions for the pion-pion-photon
cross section $d\sigma_{hh\gamma^*}$ of Eq.\ (\ref{brv}) based on the
factorization into a current and hadronic matrix element of Section
\ref{sec:external}. These approximations are based on an angular
average of $(-J^2)$ in the $ab$ center-of-momentum frame. Using
three-space invariants, we integrate the current covariantly in Section
\ref{sec:xs3sp}.

\subsection{The R\"uckl formula}
\label{sec:ruckl}

In 1976, R\"uckl proposed \cite{Ruc76a} that the dilepton cross section
be written in terms of the {\it real photon} cross section:
\begin{equation}
\label{rra}
E_+ E_-
{d\sigma_{hh\ell^+\ell^-} \over d^3\bbox{p}_+ d^3\bbox{p}_- }
=
{\alpha \over 2 \pi^2 M^2} q_0
{{d\sigma_{hh\gamma}} \over {d^3\bbox{q} }} \,.
\end{equation}
This ansatz, now in common use, was utilized by Haglin et al.
\cite{Hag93a} and others previously
\cite{Gal87a,Cle91a,Hag92a,Cra78a}
to derive dilepton production rates.
In our notation, the R\"uckl formula reads
\begin{equation}
\label{rrb}
d\sigma_{hh\ell^+\ell^-}
= {\alpha \over \pi^2 M^2} { d\tau_+ \, d\tau_- \over d\tau_q}
   d\sigma_{hh\gamma} \,,
\end{equation}
which becomes, using the formulas of the previous section,
and approximating $\int d\Omega_+ = 4\pi$
and  $\kappa \simeq \alpha / 3 \pi$,
\begin{equation}
\label{rrc}
d\sigma_{hh\ell^+\ell^-}
= {3\over 2} \, \kappa \, d\sigma_{hh\gamma} \, {dM^2 \over M^2} \,.
\end{equation}
The real-photon cross section $d\sigma_{hh\gamma}$
can be written down immediately from the virtual-photon cross section
$d\sigma_{hh\gamma^*}$ of (\ref{brv}),
\begin{equation}
\label{rrd}
d\sigma_{hh\gamma} = 4 \pi\alpha\,
(-J^2) |{\cal M}_h|^2 \, {dR_3 \over (2\pi)^5 F} \,.
\end{equation}
Neglecting $q$ in the delta function of $dR_3$, the 3-phase space integral
of Eq.\ (\ref{brw}) factorizes into the usual 2-phase space
and the $\bbox{q}$-integration
\begin{equation}
\label{rre}
dR_3 \simeq dR_2 d\tau_q \,,
\end{equation}
with
\begin{equation}
\label{rrf}
dR_2 = \delta(p_a + p_b - p_1 - p_2) \, d\tau_1 \,  d\tau_2 \,,
\end{equation}
leading all in all to
\begin{equation}
\label{rrg}
{ d\sigma_{hh\ell^+\ell^-}^{\rm Ruckl} \over dM^2 }
=
{3\over 2} \, {\kappa \over M^2} {\alpha \over \pi}
\int d\sigma_{hh} \, (-J^2) \, {d\tau_q \over 2 \pi}  \,,
\end{equation}
where
\begin{equation}
\label{rrff}
d\sigma_{hh} =  |{\cal M}_h |^2 {dR_2 \over (2\pi)^2 F }
\end{equation}
is the elastic hadronic cross section.
Integrating $(-J^2)$ first with respect to the photon phase space
with $d\tau_q = d\Omega_q |\bbox{q}| dq_0/2$ leads to
\begin{equation}
\label{rrh}
\int (-J^2) {d\tau_q \over 2 \pi}
= \int_M^A dq_0\sqrt{q_0^2 - M^2}
\int {d\Omega_q \over 4 \pi} (-J^2) \,,
\end{equation}
where $A = [s + M^2 - (m_1+m_2)^2]/2\sqrt{s}$ is the upper
kinematic boundary.

The angular average over $d\Omega_q$ was found by Haglin et
al.\ for the real-photon current (\ref{brl}) in terms
of velocities $\bbox{\beta}_i$ in the $(\bbox{p}_a + \bbox{p}_b = 0$)
cms as \cite{Hag93a}
\begin{eqnarray}
\label{hge}
\langle -I^2 \rangle &=& \int {d\Omega_q \over 4\pi} (-I^2)
\nonumber\\
&=&
{1\over q_0^2}
\left[ - (Q_a^2 + Q_b^2 + Q_1^2 + Q_2^2)
 -  2 Q_a Q_b  {\cal F}_{ab} -  2 Q_1 Q_2  {\cal F}_{12}
 + 2 \sum_{i=a,b}\sum_{j=1,2} Q_i Q_j {\cal F}_{ij}
\right] \,,
\end{eqnarray}
where
\begin{eqnarray}
\label{hgf}
{\cal F}_{ij} &\equiv&
{(1 - \bbox{\beta}_i \cdot \bbox{\beta}_j) \over  2 D_{ij} }
\left[
 \ln\left|
  \bbox{\beta}_i {\bbox{\cdot}}
 (\bbox{\beta}_j - \bbox{\beta}_i) - D_{ij}
           \over
  \bbox{\beta}_i {\bbox{\cdot}}
 (\bbox{\beta}_j - \bbox{\beta}_i) + D_{ij}
   \right|
+
(i \leftrightarrow j)
\right] \,,
\\
D_{ij} &\equiv&
       \left[   (\bbox{\beta}_i -        \bbox{\beta}_j)^2
              - (\bbox{\beta}_i {\times} \bbox{\beta}_j)^2
       \right]^{1/2} \,.
\end{eqnarray}
To convert these angular averages into functions of 2-phase space
invariants $(s,t)$, it is necessary to invoke approximation (C)
restricting phase space: While $\bbox{p}_a + \bbox{p}_b = 0$
in the chosen frame, the presence of $\bbox{q}$ means
that $\bbox{p}_1 + \bbox{p}_2 \ne 0$, i.e.\ $\bbox{\beta}_1$ and
$\bbox{\beta}_2$ are not true cms velocities. Ignoring therefore $\bbox{q}$
at this point, the velocities and hence the angular averages can be
written \cite{Hag93a} in terms of $s$, $t$ and $q_0$,
\begin{eqnarray}
\label{hgk}
\langle -I^2\rangle &=& {1\over q_0^2}
\Biggl[ - (Q_a^2 + Q_b^2 + Q_1^2 + Q_2^2)
\nonumber\\
&-&        2(Q_a Q_b + Q_1 Q_2)
         {s-2m^2\over \sqrt{s(s-4m^2)}}
       \ln\left|s+\sqrt{s(s-4m^2)} \over
                s-\sqrt{s(s-4m^2)} \right|
\nonumber\\
&+&        2(Q_a Q_1 + Q_b Q_2)
         {2m^2-t\over \sqrt{-t(s+u)}}
       \ln\left|\sqrt{-t(s+u)} - t \over
                \sqrt{-t(s+u)} + t \right|
\nonumber\\
&+&
\left.
           2(Q_a Q_2 + Q_b Q_1)
         {2m^2-u\over \sqrt{-u(s+t)}}
       \ln\left|\sqrt{-u(s+t)} - u \over
                \sqrt{-u(s+t)} + u \right|
\right] \,,
\end{eqnarray}
with $u = 4m^2 - s - t$ as usual.

Having expediently eliminated $q$ at two points, it is very important
in this formulation to correct for phase space, i.e.\ to alleviate the
effects of approximation (C). Pure phase space integration of $dR_3$
({\it including} $q$ in the delta function) yields exactly
\begin{equation}
\label{icg}
\int dR_3 =
\int d\tau_q
{ \pi\,\lambda^{1/2}(s^{\prime},m_a^2,m_b^2) \over 2s^{\prime} }
\end{equation}
where
\begin{equation}
\label{ich}
s^{\prime} = s - 2q_0 \sqrt{s} + M^2 \,.
\end{equation}
Integration over pure two-particle phase space, on the other hand,
yields
\begin{equation}
\label{ici}
\int dR_2 = { \pi\,\lambda^{1/2}(s,m_1^2,m_2^2) \over 2s } \,,
\end{equation}
so $\int dR_2$, unconstrained by $q$, grossly overestimates $\int dR_3$,
making it necessary to insert {\it a posteriori\/} the factor \cite{Gal89a}
\begin{equation}
\label{icj}
C(s,q_0) =
{s          \;  \lambda^{1/2}(s^{\prime},m_a^2,m_b^2) \over
 s^{\prime} \;  \lambda^{1/2}(s,m_1^2,m_2^2) }
\end{equation}
into the $q_0$-integration (\ref{rrh}). The ``R\"uckl approximation'' for
dilepton production
therefore reads, in its final form,
\begin{equation}
\label{ick}
{d\sigma_{hh\ell^+ \ell^-}^{\rm Ruckl}(s)
\over dM^2\ \ \ } =
{3\over 2} \, {\kappa \over M^2} \,
{\alpha \over \pi} \int d\sigma_{hh}(s,t) \,
\int dq_0 \, |\bbox{q}| \, C(s,q_0)\, \langle -I^2(s,t) \rangle \,;
\end{equation}
this is the form used in Ref.\ \cite{Hag93a}.

\subsection{Current formulae based on Lichard}
\label{sec:lichang}

Starting not from the R\"uckl formula but rather from Lichard's
improved formalism \cite{Lic95a} of Section \ref{sec:external}, the
derivation proceeds along the same lines as the above. One finds
correspondingly for the real-photon current
\begin{equation}
\label{icklz}
{d\sigma_{hh\ell^+ \ell^-}^{\rm L0}(s) \over dM^2\ \ \ } =
{\kappa \over M^2} \,
{\alpha \over \pi} \int d\sigma_{hh}(s,t) \,
\int dq_0 \, |\bbox{q}| \, C(s,q_0)\, \langle -I^2(s,t) \rangle \,,
\end{equation}
differing from the above by a constant factor $3/2$ and the extra factors
entering the exact version (\ref{bru}) of $\kappa$.

The corresponding virtual-photon current yields the third
approximation within the angle-averaged current family,
\begin{equation}
\label{icklo}
{d\sigma_{hh\ell^+ \ell^-}^{\rm L1}(s) \over dM^2\ \ \ } =
{\kappa \over M^2} \,
{\alpha \over \pi} \int d\sigma_{hh}(s,t) \,
\int dq_0 \, |\bbox{q}| \, C(s,q_0)\, \langle -J^2(s,t) \rangle \,.
\end{equation}
where the angular average of (\ref{brk}) is
\begin{eqnarray}
\label{hgh}
\langle -J^2\rangle &=&
\int {d\Omega_q \over 4\pi} (-J^2)
\nonumber\\
={}
&-&
{ (Q_a^2 + Q_b^2)(4m^2 - M^2) \over
 q_0^2 s - \bbox{q}^2 (s-4m^2) - 2q_0\sqrt{s} M^2 + M^4 }
\nonumber\\
&-&
{ (Q_1^2 + Q_2^2)(4m^2 - M^2) \over
 q_0^2 s - \bbox{q}^2 (s-4m^2) + 2q_0\sqrt{s} M^2 + M^4 }
\nonumber\\
&-&
{ Q_a Q_b (2s - 4m^2 - M^2) \over
  |\bbox{q}| v_- \sqrt{s-4m^2} }
\ln\left| |\bbox{q}| \sqrt{s-4m^2} + q_0 \sqrt{s} - M^2  \over
          |\bbox{q}| \sqrt{s-4m^2} - q_0 \sqrt{s} + M^2
  \right|
\nonumber\\
&-&
{ Q_1 Q_2 (2s - 4m^2 - M^2) \over
  |\bbox{q}| v_+ \sqrt{s-4m^2} }
\ln\left| |\bbox{q}| \sqrt{s-4m^2} + q_0 \sqrt{s} + M^2  \over
          |\bbox{q}| \sqrt{s-4m^2} - q_0 \sqrt{s} - M^2
  \right|
\\
&+&
{ (Q_a Q_1 + Q_b Q_2)(4m^2 - 2t + M^2) \over 2\chi_t }
\ln\left| (  \bbox{q}^2 t - \chi_t - M^2 v_-)
          (  \bbox{q}^2 t - \chi_t + M^2 v_+)   \over
          (  \bbox{q}^2 t + \chi_t - M^2 v_-)
          (  \bbox{q}^2 t + \chi_t + M^2 v_+)
  \right|
\nonumber\\
&+&
{ (Q_a Q_2 + Q_b Q_1)(4m^2 - 2u + M^2) \over 2\chi_u }
\ln\left| (  \bbox{q}^2 u - \chi_u - M^2 v_-)
          (  \bbox{q}^2 u - \chi_u + M^2 v_+)   \over
          (  \bbox{q}^2 u + \chi_u - M^2 v_-)
          (  \bbox{q}^2 u + \chi_u + M^2 v_+)
  \right| ,
\nonumber
\end{eqnarray}
where
\begin{eqnarray}
\label{hgi}
v_{\pm} &\equiv& q_0\sqrt{s} \pm M^2 \,, \\
\label{hgj}
\chi_t &\equiv&
|\bbox{q}| \sqrt{-t(s q_0^2 + u \bbox{q}^2) - u M^4} \,,\\
\label{hgak}
\chi_u &\equiv&
|\bbox{q}| \sqrt{-u(s q_0^2 + t \bbox{q}^2) - t M^4} \,.
\end{eqnarray}
Eq.\ (\ref{hgh}) reduces to Eq.\ (\ref{hgk}) for $M\to 0$.  We shall be
testing the effect of using the virtual over real-photon currents
within the present approximation in Section \ref{sec:results} below.
We note in passing that Eq.\ (\ref{hgh}) would be preferable to
Eq.\ (\ref{hgk}) for very small hadron masses $m$, since then the
latter is logarithmically sensitive to $1/m$ while the former is
regulated by the dilepton invariant mass $M$.

\subsection{Three-phase space current approximations}
\label{sec:xs3sp}

Three approximations were needed for the above angle-averaged current
cross sections:  the R\"uckl formula, converting cms velocities
$\bbox{\beta}_i$ to invariants $s$ and $t$ leading to Eqs.\ (\ref{hge})
and (\ref{hgf}), and correcting for phase space using
Eq.\ (\ref{icj}).  At the cost of increased computing time, all three
can be avoided by utilizing the formalism of three-particle phase space
invariants.

Within 3-phase space, and writing the flux as $F = 2 \lambda^{1/2}
(s,m_a^2,m_b^2)$, the bremsstrahlung cross section of Eq.\
(\ref{brv}) becomes\footnote{
Technically, it would in equation (\ref{bvp}) be possible to restore
$q$ to ${\cal M}_h$ and then to include it with $-J^2$ in the inner
integral, but because of missing contact terms this procedure is not
gauge invariant and shall not be pursued.}
\begin{equation}
\label{bvp}
{d\sigma_{hh\ell^+ \ell^-}\over dM^2}(s) \simeq
{4 \pi\alpha \over (2\pi)^5 M^2}  \,
{\kappa(M^2)\pi \over 8 \lambda(s,m_a^2,m_b^2)}
\int dt_1\, |{\cal M}_h(s,t_1)|^2
\int { ds_2\, ds_1\, dt_2 \over  \sqrt{B}}
\; \left[-J^2(s,t_1,s_2,s_1,t_2)\right]  \,.
\end{equation}
This differs from Eq.\ (\ref{icklo}) in the way the current is treated:
by an angular average plus $q_0$-integration for the earlier, by full
3-phase space invariants in the present case. Note that ${\cal M}_h$ is
here a function of $t_1$ rather than $t$.

\section{Exact cross section}
\label{sec:xsexact}

In all the above approximations, neglecting the dependence on $q$ of
the squared hadronic matrix element $|{\cal M}_h(p_a,p_b,p_1,p_2)|^2$
is unavoidable because this had been a precondition to factorizing into
electromagnetic and hadronic parts in Eq.\ (\ref{bre}).  There have
been attempts go beyond this on-shell matrix element approximation by
expanding ${\cal M}_h$ in $q$
(see Refs.\ \cite{Lic95a,Zha95a,Cra78a}),
but this procedure involves successively more
complicated derivatives without guaranteeing convergence.

Using 3-phase space invariants and the exact\footnote{
To order $\alpha$
for the emission of real photons and to order $\alpha^2$ for dilepton
emission.}
expressions (\ref{fcm}) for the pion-pion-photon cross section and
(\ref{fcl}) for the dilepton cross section, such expansions in $q$
become superfluous.  To calculate explicit dilepton cross sections, we
merely insert from our favorite model for pion-pion scattering the
complete set of matrix elements, and write the elements of the squared
sum of (\ref{fcm}) in terms of these invariants:
\begin{equation}
\label{xxb}
{\cal M}_m^\mu {\cal M}_{n\mu}^*(p_a,p_b,p_1,p_2,q)
\longrightarrow
{\cal M}_m^\mu {\cal M}_{n\mu}^*(s,t_1,s_2,s_1,t_2)
\equiv T_{mn}  \,.
\end{equation}
This casting in terms of invariants means that $q$ and all
consequent off-shell effects in $|{\cal M}|^2$ are fully
taken care of.\footnote{
Matrix elements keeping their full dependence on $q$ can be implemented
within 2-phase space when only the {\it differential} cross section
$d\sigma/d^3\bbox{q}\, dM^2$ need be calculated; see e.g.\ Ref.\
\cite{Hag91a}; for $d\sigma/dM^2$, this becomes impossible. }

Putting it all together, we obtain
\begin{equation}
\label{xxc}
{d\sigma_{hh\ell^+ \ell^-}^{\rm exact} \over dM^2}(s) =
{4 \pi\alpha \over (2\pi)^5 M^2}  \,
{\kappa(M^2)\pi \over 8 \lambda(s,m_a^2,m_b^2)}
\int { dt_1\,  ds_2\, ds_1\, dt_2 \over  \sqrt{B}}
\left[-\sum_{mn} T_{mn}(s,t_1,s_2,s_1,t_2)\right]  \,.
\end{equation}
While the cross sections may be exact, microscopic models for the
pion-pion interaction are far from perfect. In the next section, we
make use of a simple model to illustrate the present cross section and
the approximations enumerated before.  The main point of this section,
however, is to stress that, apart from model-induced ailments, the
exact treatment of any model for the reaction $\pi\pi \to
\pi\pi\ell^+\ell^-$ is possible.

\section{Gauge-invariant OBE model}
\label{sec:fullobe}

To compare dilepton cross sections under the exact formula and various
approximations introduced so far, we must make a model: little if any
experimental data for the differential pion-pion-photon cross section
is available. To this purpose, we return to a simple lagrangian with
$\sigma$, $\rho$ and $f(1270)$ exchange used previously \cite{Hag93a}
which, when $t$-integrated, successfully reproduced experimental data
for elastic $\pi^+ \pi^- \to \pi^+ \pi^-$ scattering
\cite{Pro73a,Sri76a}:
\begin{equation}
\label{rtb}
{\cal L}_{\rm int} =
  g_\sigma\, \sigma
   \partial_\mu \bbox{\pi {\cdot}} \partial^\mu \bbox{\pi}
+ g_\rho  \, \rho^\mu
   \bbox{\pi {\times}} \partial_\mu \bbox{\pi}
+ g_f \, f_{\mu\nu}
   \partial^\mu \bbox{\pi {\cdot}} \partial^\nu \bbox{\pi} \,.
\end{equation}
Here, we include the full momentum dependence of the $f$ propagator
which now reads \cite{Vel70a,Vel76a}
\begin{eqnarray}
\label{rtc}
i{\cal P}_{\alpha\beta\gamma\delta}(k)
&=& { -i\; f_{\alpha\beta\gamma\delta}(k)
\over k^2 - m_f^2 + i m_f \Gamma_f} \,,\\
\label{rtd}
f_{\alpha\beta\gamma\delta}(k)
&=&
{1\over 2}
\left( g_{\alpha\gamma} g_{\beta\delta}
     + g_{\alpha\delta} g_{\beta\gamma}
     - g_{\alpha\beta}  g_{\gamma\delta}
\right)
\nonumber\\
&-&
{1\over 2}
\left( g_{\alpha\gamma} { k_\beta  k_\delta \over m_f^2 }
     + g_{\alpha\delta} { k_\beta  k_\gamma \over m_f^2 }
     + g_{\beta\gamma}  { k_\alpha k_\delta \over m_f^2 }
     + g_{\beta\delta}  { k_\alpha k_\gamma \over m_f^2 }
\right)
\nonumber\\
&+&
{2\over 3}
\left( {1\over 2} g_{\alpha\beta}  + {k_\alpha k_\beta \over m_f^2}\right)
\left( {1\over 2} g_{\gamma\delta} + {k_\gamma k_\delta\over m_f^2}\right)
\,.
\end{eqnarray}
The $\bbox{\rho}$ field also has a $k$-dependent term in the
unitary gauge propagator \cite{Ser86a},
\begin{equation}
\label{rtg}
iR_{\mu\nu}(k) =
{i \over k^2 - m_\rho^2 + i m_\rho \Gamma_\rho}
\left[
-g_{\mu\nu} + {k_\mu k_\nu \over  m_\rho^2} \right]
\,,
\end{equation}
but this term contributes to matrix elements only in the form
$(m_a^2-m_1^2)(m_b^2-m_2^2)$ etc.\ and is thus zero for equal pion
masses.

For $t$- and $u$-channel exchange of an $\alpha$-meson ($\alpha =
\sigma,\ \rho\ f$) with momentum $k$ by the pions, we implement
monopole strong form factors \cite{Ris84a}
\begin{equation}
\label{rth}
h_\alpha(k) = {m_\alpha^2 - m_\pi^2 \over m_\alpha^2 - k^2} \,.
\end{equation}
In Figure 4, we show the results of fitting this model to the elastic
scattering data as before \cite{Hag93a} but with the different
propagator (\ref{rtd}).  Again, the fit is reasonably good, but the
strong dependence on momentum-dependent terms in the $f$ propagator
shows up in the region above the $f$ peak. Best-fit values for the free
parameters now are: $m_\sigma = 475$ MeV, $m_\rho = 775$ MeV, $m_f =
1220$ MeV; $m_\sigma g_\sigma = 3.1$, $g_\rho = 6.15$, $m_f g_f =
8.85$.  Throughout, we take $m_\pi = 140$ MeV.

For the current-based approximations of Section \ref{sec:curint}, the
above lagrangian is all we need; for the complete cross section of
Section \ref{sec:xsexact}, it must be augmented by internal photon
emission pieces.  This can be achieved by minimal substitution, thereby
guaranteeing gauge invariance (for a similar approach, see the
``Quantum Hadrodynamics'' lagrangians of Ref.\ \cite{Ser86a}).  To the
hadronic lagrangian (\ref{rtb}) we therefore add
\begin{equation}
\label{fub}
{\cal L}_{\rm em} =
  {\cal L}_{\pi\pi\gamma}
+ {\cal L}_{\bbox{\rho\rho}\gamma}
+ {\cal L}_{\pi\pi\sigma\gamma}
+ {\cal L}_{\pi\pi\bbox{\rho}\gamma}
+ {\cal L}_{\pi\pi f\gamma} \,.
\end{equation}
Details regarding these pieces and corresponding vertex factors
can be found in the Appendix.

The $\rho\rho\gamma$ interaction results in a modified
strong form factor for $t$- or $u$-channel charged $\rho$ exchange.
If $h_k$ and $h_l$ are the usual form factors at the
$\pi\pi\rho(k)$ and $\pi\pi\rho(l)$ vertices respectively, with
$k = l + q$, electromagnetic gauge invariance requires that the
form factor for the emission of a photon by the $\rho$ be
\begin{equation}
\label{lld}
H(k,l) = {u h^2_l - v h^2_k \over (u-v) h_k h_l} \,,
\end{equation}
where
$u = k^2 - m_\rho^2 + i m_\rho \Gamma_\rho$ and
$v = l^2 - m_\rho^2 + i m_\rho \Gamma_\rho$
are the respective $\rho$ propagators. For the monopole form
(\ref{rth}) of $h$ this becomes \cite{Ris84a,Tow87a}
\begin{equation}
\label{lle}
H(k,l) = 1
+ \left({m_\rho^2 - k^2 - i m_\rho \Gamma_\rho
    \over m_\rho^2 - l^2}\right)
+ \left({m_\rho^2 - l^2 - i m_\rho \Gamma_\rho
    \over m_\rho^2 - k^2}\right) \,.
\end{equation}
Note that we do not include strict Vector
Meson Dominance coupling of the photons to vector mesons. This is
equivalent to the introduction of electromagnetic form factors. For the
specific application we are considering, small $M$'s, those would be
$\approx$ 1. For previous OBE models, see e.g.\ Refs.\
\cite{Hag91a,Schae94a} and references therein.

\section{Results and conclusions}
\label{sec:results}

To summarize: the differential cross section for dilepton production as
a function of invariant mass, $d\sigma_{hh\ell^+\ell^-}/dM$, can be
found in a number of approximations and an exact way (within the limitations
of the hadronic reaction model). The approximations are:  the R\"uckl
approximation (\ref{ick}), two approximations, Eqs.\ (\ref{icklz}) and
(\ref{icklo}), based on a more careful derivation by Lichard, using the
real and virtual photon currents respectively, and an approximation
integrating the current covariantly in 3-phase space, Eq.\ (\ref{bvp}).
The ``exact'' formula is given by Eq.\ (\ref{xxc}).

Of these, the R\"uckl and Lichard cross sections can be found easily
and quickly within 2-phase space. The 3-phase space current
approximation is somewhat more difficult but still uses 2-phase
space hadronic matrix elements. Implementing the exact formulation
(\ref{xxc}) involves much more effort: if there are $N$
diagrams contributing to a given reaction, the sum in (\ref{xxc})
contains $N(N+1)/2$ terms. Since an individual term $T_{mn}$ in
(\ref{xxb}) is itself a Dirac sum, only partial factorization
of the overall sum is possible. Further complications arise from
the presence of imaginary pieces in the matrix elements.

To quantify the differences between the approximations and the exact
formulation, we have studied all five distinct pion-pion reactions.
Writing $({+}{-}) \to ({+}{-})$ as shorthand for the reaction
$(\pi^+\pi^- \to \pi^+\pi^- \ell^+ \ell^-)$ and so on, we have
calculated within our OBE model cross sections for
$({+}{-}) \to ({+}{-})$,
$({+}{+}) \to ({+}{+})$,
$({+}{-}) \to ({0}{0})$,
$({0}{0}) \to ({+}{-})$ and
$({+}{0}) \to ({+}{0})$.
A total of 36 diagrams contribute at tree level to the first two reactions
(24 for $\gamma^*$ emitted by external pion lines, 12 by vertices
and the exchange meson); the remaining reactions are made up of
16 diagrams each (8 external, 8 internal).

Numerical results were checked by performing the following consistency
checks:  setting $s_1 \to s$, $t_1 \to t$, $t_2 \to m_\pi^2$, $s_2 \to
m_\pi^2$ and $M^2 \to 0$ in the {\it hadronic} part of the exact matrix
elements (while keeping these variables where they enter the
``current'' part) and using only the external-emission diagrams
reproduces exactly the results of the 3-phase space integrated current
(\ref{bvp}).  Gauge invariance in the $\sigma$, $\rho$ and $f$ fields
provided another sensitive test.

Figures 5--9 show the cross sections for these five reactions as a
function of $\sqrt{s}$. In every case, the left and right panels show
$d\sigma/dM$ for $M = 10$ MeV and 300 MeV respectively.  Final-state
symmetrization factors were included where appropriate. Initial-state
symmetrization was also included for $({+}{+}) \to ({+}{+})$ and
$({0}{0}) \to ({+}{-})$ in order to facilitate  use within a thermal
pion gas environment. When the initial-state pions are considered
distinguishable, these two cross sections should be multiplied by a
factor 2 for these reactions.  Because they are identical in structure
to their charge-conjugate versions, cross sections for the reactions
$({+}{0}) \to ({+}{0})$ and $({+}{+}) \to ({+}{+})$ were doubled ---
this, too, should be corrected for when desired.

To prevent overcrowding, only the R\"uckl plus the virtual-$\gamma^*$
current approximations (\ref{icklo}) and (\ref{bvp}) are shown,
together with the exact calculations. All these were computed using the
same OBE model and parameter values specified in Section
\ref{sec:fullobe} above.

In the complicated structures and deviations in Figs.\ 5--9, the
following points are of interest:
\begin{enumerate}
  \item
The $\rho$ peak in the reaction $({+}{-}) \to ({+}{-})$
is overestimated by a factor 2.5 and 1.5
by the R\"uckl and Lichard approximations respectively for
$M = 10$ MeV; for $M = 300$ MeV, the overestimation is much greater
(Figure 5).
Only the 3-space current approximation does an adequate job here.
Similarly, the $f$ peak of the exact result lies well below
the corresponding approximations. Overestimation (1.7--5.0) of the $\rho$
peak also occurs for approximations in $({+}{0}) \to ({+}{0})$.
  \item
In the reaction $({+}{0}) \to ({+}{0})$, the 3-space current
approximation, on the other hand, lies well below the exact result
(Figure 6).
  \item
Overestimation factors for $d\sigma/dM$ are of order 0.5--4 for the
reactions $({0}{0}) \to ({+}{-})$ and $({+}{-}) \to ({0}{0})$,
depending on approximation and cms energy. Here, too, the
3-space current approximation generally does better than the others
but again tends to underestimate in some parts (Figs.\ 7, 8).
   \item
The largest discrepancy between approximations and the exact result
occur for the reaction $({+}{+}) \to ({+}{+})$ (Figure 9):
for the R\"uckl approximation, factors 3 (for 10 MeV) to
30 (for 300 MeV) arise, while the Lichard approximation yields
corresponding overestimation factors of 1.9 and 14--20.
Since, however, this reaction contributes only little when
contributions from all reactions are added up, this effect is not
of much interest in a heavy-ion context.
   \item
There is generally not much difference between using the real-photon
(not shown) or virtual-photon currents in the Lichard approximations;
this is in agreement with the conclusions of Ref.\ \cite{Zha95a}.
For $M = 300$ MeV, some differences can be discerned, but both deviate
in general more from the exact result than from each other.
\end{enumerate}

Figures 10 and 11 show the dilepton production rates per unit spacetime
summed over all seven reactions, calculated using the Boltzmann formula
(\ref{rtk}), for temperatures $T = 100$ and 200 MeV respectively. These
temperatures probably lie below and above an expected transition to a
new quark-gluon phase and so represent extrema in their behavior.
Corresponding ratios of approximation over exact rates are shown in
Figures 12 and 13.  The R\"uckl approximation is the worst, as expected;
overestimation is in the range 2--4 for $T = 100$ MeV, and 2--8 for $T
= 200$ MeV.  The Lichard approximations overestimate by factors 1.4--4,
depending on temperature and $s$.  Here, the {\it real-$\gamma$}
3-phase space current, the lower of the two dotted lines, does
surprisingly well, deviating from the exact result by less than 20\%
throughout. Comparing this fact, however, with the substantial
deviations shown in the $s$-differential cross sections of Figs.\ 5--9,
one must conclude that much of the ``agreement'' seen on the level of
rates is due to averaging over differences seen in the $s$-dependent
cross sections and that such ``agreement'' therefore does not validate
the approximations.

We also note the fact that none of the approximations approaches the
exact result for small values of $M$: even for the smallest value shown
($M = 10$ MeV), the discrepancy is still above 40\% for the Lichard
approximations and larger than a factor 2 for the R\"uckl
approximation.

Figure 14 shows the fractional contributions to the total $T = 200$ MeV
rate for the R\"uckl, virtual-$\gamma$ current Lichard, and exact cross
sections. As discussed, the $({+}{+}) \to ({+}{+})$ contribution
virtually falls away in the exact calculation; by contrast the reaction
$({+}{0}) \to ({+}{0})$ plus charge conjugate becomes more important,
in second place behind the dominant reaction $({+}{-}) \to ({+}{-})$.

Finally, we ask where the discrepancies between approximations and
exact cross section originate: is this due to the inclusion of diagrams
with photon emission from vertices and propagators? A partial answer is
provided by Figure 15, where the cross sections for reactions $({+}{+})
\to ({+}{+})$ and $({+}{-}) \to ({0}{0})$ are plotted. The upper lines
correspond, as before, to the R\"uckl and two Lichard approximations,
while the solid line again represents the exact result.  The lowest
dash-dotted line, on the other hand, represents the exact result but
excluding all internal diagrams and their cross terms with external
ones. The difference between this lower line and the exact result
(solid line) therefore represents the contribution of the internal
diagrams; while the difference between the lower dash-dotted line and
the upper lines (approximations) represents the change in cross section
due to inclusion/exclusion of $q$ in the {\it external}-emission
diagrams.

We see that the contribution of internal emission is not all that
large, albeit nonnegligible. By far the most important effect on
$d\sigma/dM$ is the inclusion of the full dependence of ${\cal M}$
on the photon momentum $q$. In fact,
the effect is so large that for the reaction $({+}{+}) \to ({+}{+})$
the cross section would become negative\footnote{
Since this curve does
not represent a gauge-invariant calculation, this negative cross
section is of no physical consequence. In fact, it illustrates
once again the necessity of including internal diagrams to make
the cross section positive again.}
if internal diagrams were not included!
In other words, among the list of approximations listed in
Section \ref{sec:spa}, approximation (B) is the most far-reaching.

The importance of including $q$ fully in the matrix element can be
understood qualitatively: as illustrated in Figs.\ 5--9 for one
reaction, the elastic pion-pion cross sections vary considerably within
the region of interest. Adding or subtracting a sizeable amount of
energy-momentum in the form of $q$ therefore naturally leads to a
substantial change in cross section also. This change
is, however, neglected in all the approximations listed\footnote{
This also explains why the overestimation is so large for the
reaction $({+}{+}) \to ({+}{+})$: there are no resonances in
the cross section which could provide some amount of cancellation
by pion-pion reactions being ``shifted into'' and ``shifted out of''
a resonance peak.}.
We believe that this (unavoidable) neglect is at the heart of the
considerable differences between approximations and exact result
seen in Figs.\ 5--9.

In summary, the various approximations for bremsstrahlung
from pion-pion collisions cannot in general be believed beyond
at best a factor 2 or more, depending on invariant mass and temperature.

We have shown in Sections \ref{sec:fullcross} and \ref{sec:xsexact} how
3-phase space can be fully taken into account. The most important
effect, it turns out, is not the inclusion of photon emission from
vertices or propagators but the strong dependence of the cross section
on the inclusion of the photon momentum $q$ into the {\it hadronic}
collision part.

Since this calculation neglected baryonic degrees of freedom and
resonances, and because of the technical difficulties of a further
integration over spacetime, our results are not applied directly
to experimental data; this is a matter for the future.  All we can say
at this stage is that it is likely that the approximations overestimate
the dilepton yield throughout the mass region of interest, and that the
degree of overestimation rises both with increasing invariant mass $M$
and temperature $T$.  Since the bremsstrahlung channel
competes with various Dalitz decays, and since various experiments take
much care in applying cuts to eliminate Dalitz pairs as far as possible
\cite{Tse95a} a quantitative estimate would necessarily have to take
such experimental cuts into account also.

In addition, attention should be paid to the issue of the
Landau-Pomeranchuk effect, which is expected to suppress dilepton
bremsstrahlung rates \cite{Cle93a,Cle93b,Kno95a}.  Our
results, taken in conjunction with these calculations, therefore appear
to indicate even larger suppression of bremsstrahlung dileptons than
had been thought previously.

\acknowledgements
We thank P.\ Lichard for enlightening discussions and his careful
derivation of dilepton formulas. Thanks to A.\ Drees, J.\ Zhang,
J.\ C.\ Pan, P.\ Lipa and B.\ Buschbeck for useful discussions.
This work was supported in part by the Natural Sciences and Engineering
Research Council of Canada, by the Qu\'ebec FCAR fund, by a NATO
Collaborative Research Grant, by the National Science Foundation under
grant number 94--03666 and by the Austrian Fonds zur F\"orderung der
wissenschaftlichen Forschung (FWF).

\appendix
\section*{Full OBE lagrangian}

Besides the hadronic interaction lagrangian (\ref{rtb}) with its
vertex factors
\begin{eqnarray}
\begin{array}{rcll}
\label{hpps}
\Gamma_{\pi\pi\sigma} &=&
\left\{
  \begin{array}{l}
    2i g_\sigma\, p\cdot p^\prime \\
     i g_\sigma\, p\cdot p^\prime  \\
  \end{array}
\right.
&
  \begin{array}{l}
    \mbox{\hspace*{4mm}for }
     \pi^{\pm}(p) \to \pi^{\pm}(p^\prime)\, \sigma \\
    \mbox{\hspace*{4mm}for }
     \pi^{0}(p) \to \pi^0(p^\prime)\, \sigma \\
  \end{array}
\\
\label{hppr}
\Gamma_{\pi\pi\rho} &=&
\left\{
  \begin{array}{l}
    \mp i g_\rho (p + p^\prime)^\mu \\
    \pm i g_\rho (p + p^\prime)^\mu \\
    \pm i g_\rho (p + p^\prime)^\mu \\
  \end{array}
\right.
&
  \begin{array}{l}
    \mbox{\hspace*{4mm}for }
     \pi^{\pm}(p) \to \pi^{\pm}(p^\prime)\, \rho_0^\mu \\
    \mbox{\hspace*{4mm}for }
     \pi^{\pm}(p) \to \pi^{0}(p^\prime)\, \rho_\pm^\mu \\
    \mbox{\hspace*{4mm}for }
     \pi^{0}(p) \to \pi^{\pm}(p^\prime)\, \rho_\mp^\mu \\
  \end{array}
\\
\label{hppf}
\Gamma_{\pi\pi f} &=&
\left\{
  \begin{array}{l}
    i g_f ( p^\alpha p^{\prime \beta}
           + p^\beta p^{\prime \alpha}) \\
    i g_f \; p^\alpha p^{\prime \beta}
  \end{array}
\right.
&
  \begin{array}{l}
    \mbox{\hspace*{4mm}for }
     \pi^{\pm}(p) \to \pi^{\pm}(p^\prime)\, f^{\alpha\beta} \\
    \mbox{\hspace*{4mm}for }
     \pi^{0}(p) \to \pi^0(p^\prime)\, f^{\alpha\beta} \\
  \end{array}
\\
\end{array}
\end{eqnarray}
we need the electromagnetic interaction lagrangian linear in $A_\mu$,
\begin{equation}
\label{ggf}
{\cal L}_{\rm em} =
  {\cal L}_{\pi\pi\gamma}
+ {\cal L}_{\bbox{\rho\rho}\gamma}
+ {\cal L}_{\pi\pi\sigma\gamma}
+ {\cal L}_{\pi\pi\bbox{\rho}\gamma}
+ {\cal L}_{\pi\pi f\gamma} \,.
\end{equation}
With the pion current
\begin{equation}
\label{ggd}
j_\pi^\mu \equiv i(  \pi^- \partial^\mu \pi^+
                   - \pi^+ \partial^\mu \pi^-) \,,
\end{equation}
the $\bbox{\rho}$ field tensor,
\begin{equation}
\label{gge}
\bbox{B}^{\mu\nu} \equiv
\partial^\mu \bbox{\rho}^\nu - \partial^\nu \bbox{\rho}^\mu
 - g_\rho \bbox{\rho}^\mu \bbox{{\times} \rho}^\nu \,,
\end{equation}
and $F_{\mu\nu} \equiv \partial_\mu A_\nu - \partial_\nu A_\mu$ the
usual electromagnetic tensor, the individual contributions are
\begin{eqnarray}
\label{ggg}
{\cal L}_{\pi\pi\gamma}
&=& - e A_\mu j_\pi^\mu \,,  \\
\label{ggh}
{\cal L}_{\bbox{\rho\rho}\gamma}
&=& e A_\mu [\bbox{\rho}_\nu \bbox{{\times} B}^{\mu\nu}]_3
   + {1\over 2}
     e F_{\mu\nu} [\bbox{\rho}^\mu \bbox{{\times} \rho}^\nu]_3
\\
&=& ie A_\mu \left[
    \rho_{- \nu} (  \partial^\mu\rho_+^\nu
                  - \partial^\nu\rho_+^\mu )
  - \rho_{+ \nu} (  \partial^\mu\rho_-^\nu
                  - \partial^\nu\rho_-^\mu )
             \right]
  + ie \partial_\mu A_\nu \left[
       \rho_-^\mu  \rho_+^\nu - \rho_-^\nu  \rho_+^\mu
                         \right]\,,
\nonumber\\
\label{ggi}
{\cal L}_{\pi\pi\sigma\gamma}
&=& - 2 e g_\sigma A_\mu j_\pi^\mu \,,  \\
\label{ggj}
{\cal L}_{\pi\pi\rho\gamma}
&=& - e g_\rho A_\mu \left[
    \bbox{\pi {\times}}(\bbox{\pi {\times} \rho}^\mu)
                    \right]_3
\\
&=& e g_\rho A_\mu
   \left[ 2 \pi^+ \pi^- \rho_0^\mu
          - \pi^- \pi^0 \rho_+^\mu
          - \pi^+ \pi^0 \rho_-^\mu
   \right]\,,
\nonumber\\
\label{ggk}
{\cal L}_{\pi\pi f\gamma}
&=& - e g_f f_{\mu\nu} \left[ A^\mu j_\pi^\nu + A^\nu j_\pi^\mu
                     \right]\,.
\end{eqnarray}
The corresponding vertex factors are
\begin{eqnarray}
\begin{array}{rcll}
\label{vtppg}
\Gamma_{\pi\pi\gamma} &=& \mp ie(p+ p^\prime)^\mu
  & \mbox{\hspace*{5mm}for }
     \pi^{\pm}(p) \to \pi^{\pm}(p^\prime) A^\mu \\
\label{vtrrg}
\Gamma_{\rho\rho\gamma} &=&
    \pm ie \left[ g^{\alpha\beta} (p+p^\prime)^\mu
            - g^{\alpha\mu} p^{\prime\beta}
            - g^{\beta\mu}  p^{\alpha}
       \right]
    & \mbox{\hspace*{5mm}for }
       \rho_\pm^\alpha(p) \to \rho_\pm^\beta(p^\prime) A^\mu(q) \\
\label{vtppsg}
\Gamma_{\pi\pi\sigma\gamma} &=&
   \pm 2ie g_\sigma (p+p^\prime)^\mu & \mbox{\hspace*{5mm}for }
                             \pi^{\pm}(p) \to
                            \pi^{\pm}(p^\prime)
                            \sigma A^\mu \\
\label{vtpprg}
\Gamma_{\pi\pi\rho\gamma} &=&
\left\{
  \begin{array}{l}
    -2ie g_\rho g^{\mu\nu} \\
    +ie g_\rho g^{\mu\nu} \\
    +ie g_\rho g^{\mu\nu} \\
  \end{array}
\right.
&
  \begin{array}{l}
    \mbox{\hspace*{4mm}for }
     \pi^{\pm}(p) \to \pi^{\pm}(p^\prime) \rho_0^\nu A^\mu \\
    \mbox{\hspace*{4mm}for }
     \pi^{\pm}(p) \to \pi^0(p^\prime) \rho_{\pm}^\nu A^\mu \\
    \mbox{\hspace*{4mm}for }
     \pi^{0}(p) \rho_{\pm}^\nu \to \pi^{\pm}(p^\prime) A^\mu \\
  \end{array}
\\
\label{vtppfg}
\Gamma_{\pi\pi f\gamma} &=&
   \pm ie g_f \left[ g^{\alpha\mu}(p+p^\prime)^\beta
                   + g^{\beta \mu}(p+p^\prime)^\alpha \right]
                          & \mbox{\hspace*{5mm}for }
                             \pi^{\pm}(p) \to
                            \pi^{\pm}(p^\prime)
                            f^{\alpha\beta} A^\mu   \,.
\end{array}
\end{eqnarray}



\vspace{9mm}

\noindent
{\Large\bf List of Figures}

\vspace{9mm}

\noindent
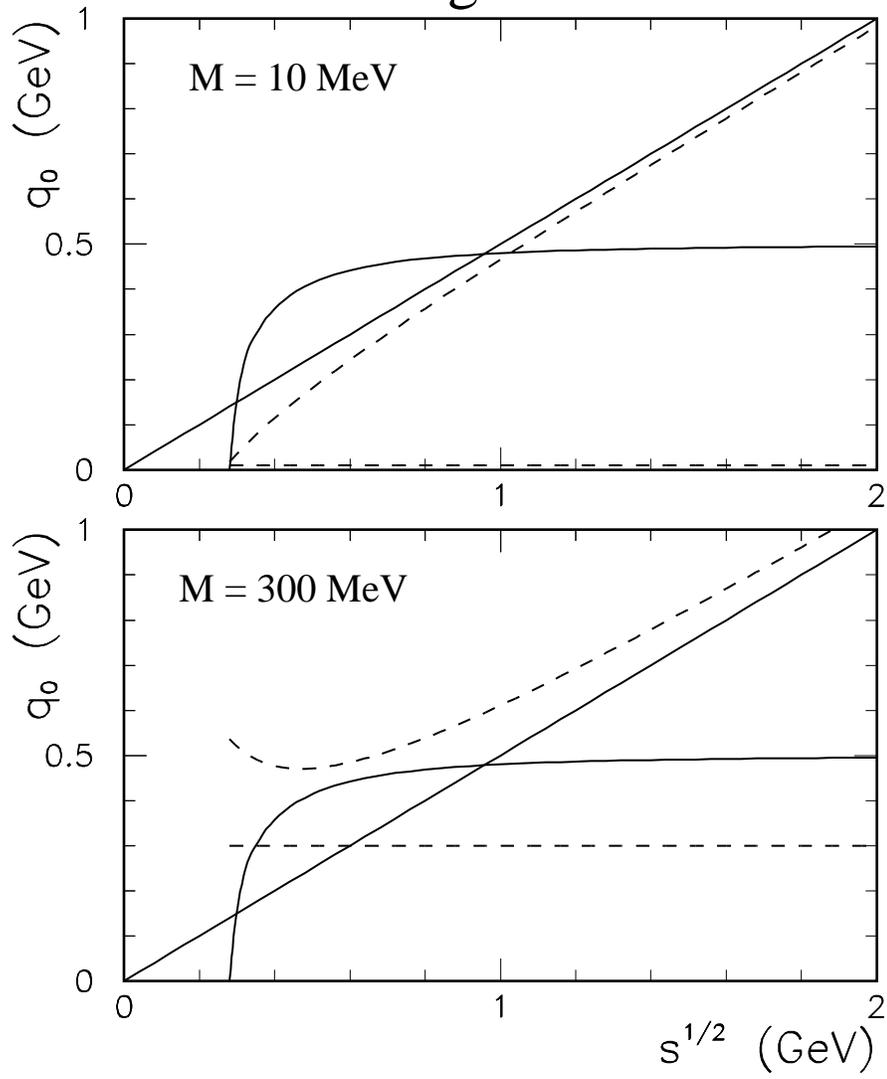
Figure 1:
Breakdown of the soft photon approximation: the SPA is valid only when
$q_0$ is much smaller than the two solid lines shown. The region between
the dashed lines is the domain of integration for $q_0$ when
calculating cross sections as a function of dilepton invariant mass.

\vspace{9mm}

\noindent
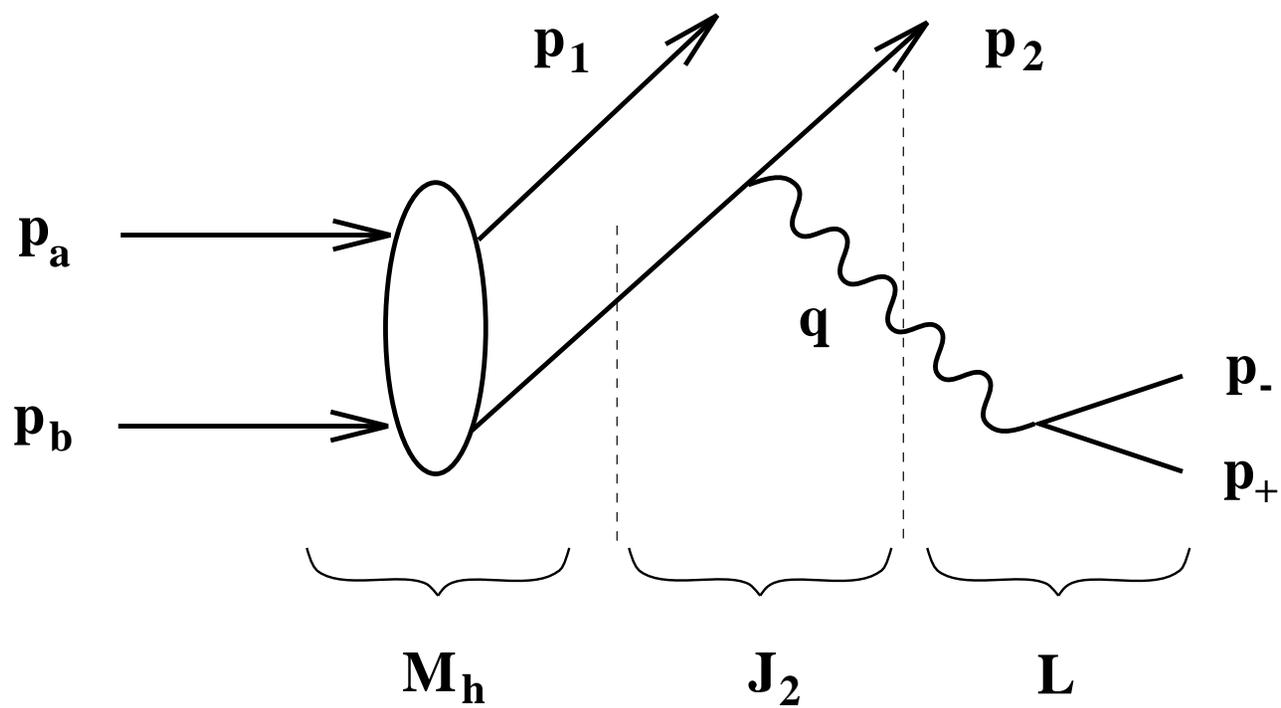
Figure 2:
One of the four contributing diagrams following from
SPA factorization of the matrix element into leptonic tensor $L_{\mu\nu}$,
photon emission current $J^\mu$, and elastic hadronic matrix element
${\cal M}_h$.

\vspace{9mm}

\noindent
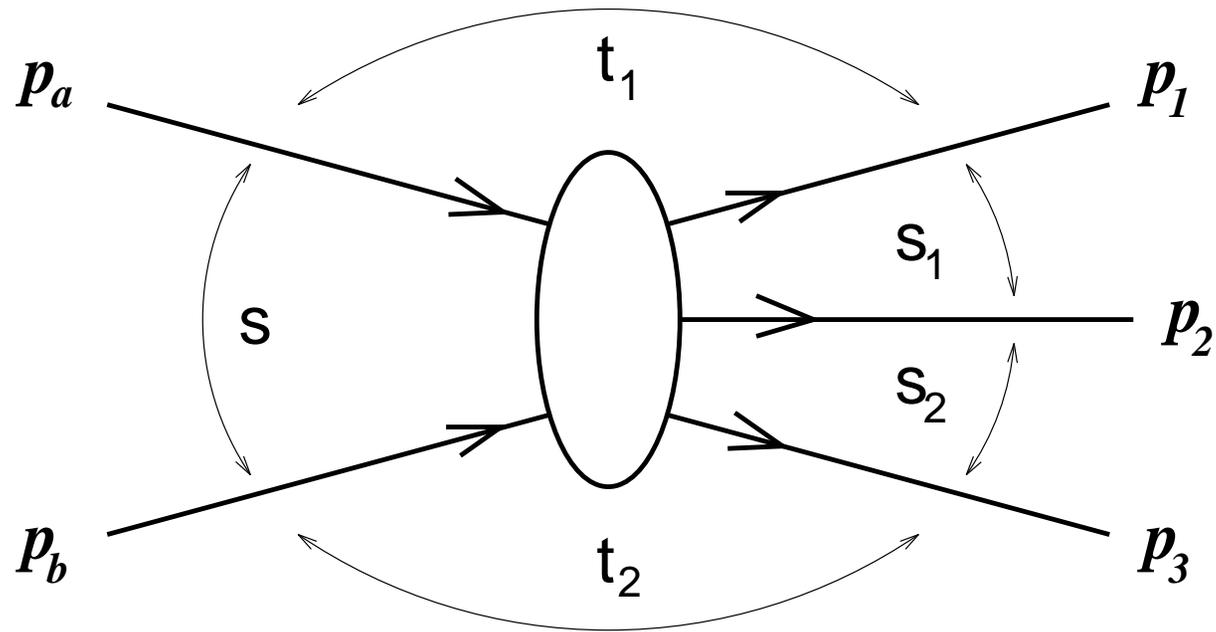
Figure 3:
Relativistic invariants for three-particle final state phase space.

\vspace{9mm}

\noindent
Figure 4:
OBE model fitted to the elastic $\pi^+\pi^- \to  \pi^+\pi^-$
cross section. Parameter values determined from this fit are used
for all reactions throughout this paper.

\vspace{9mm}

\noindent
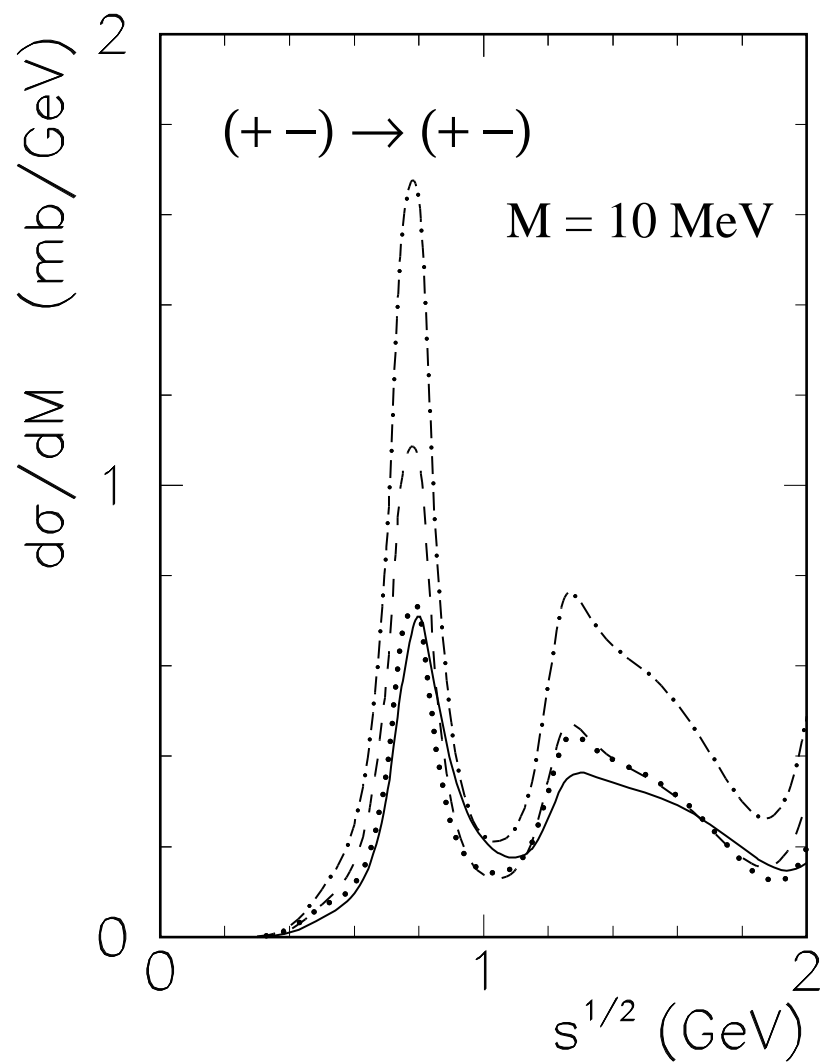
Figure 5:
Cross section for the reaction
$\pi^+\pi^- \to \pi^+\pi^- e^+ e^-$ as a function of $\sqrt{s}$
for fixed dielectron invariant masses $M = 10$ and 300 MeV.
Solid line: exact OBE calculation Eq.\ (\ref{xxc}).
Dash-dotted line: R\"uckl approximation (\ref{ick}) --- this is
the commonly-used version of the SPA. Other approximations are:
Dashed line: angular-averaged virtual $\gamma$ current (\ref{icklo}).
Dotted line: 3-phase space current (\ref{bvp}).

\vspace{9mm}

\noindent
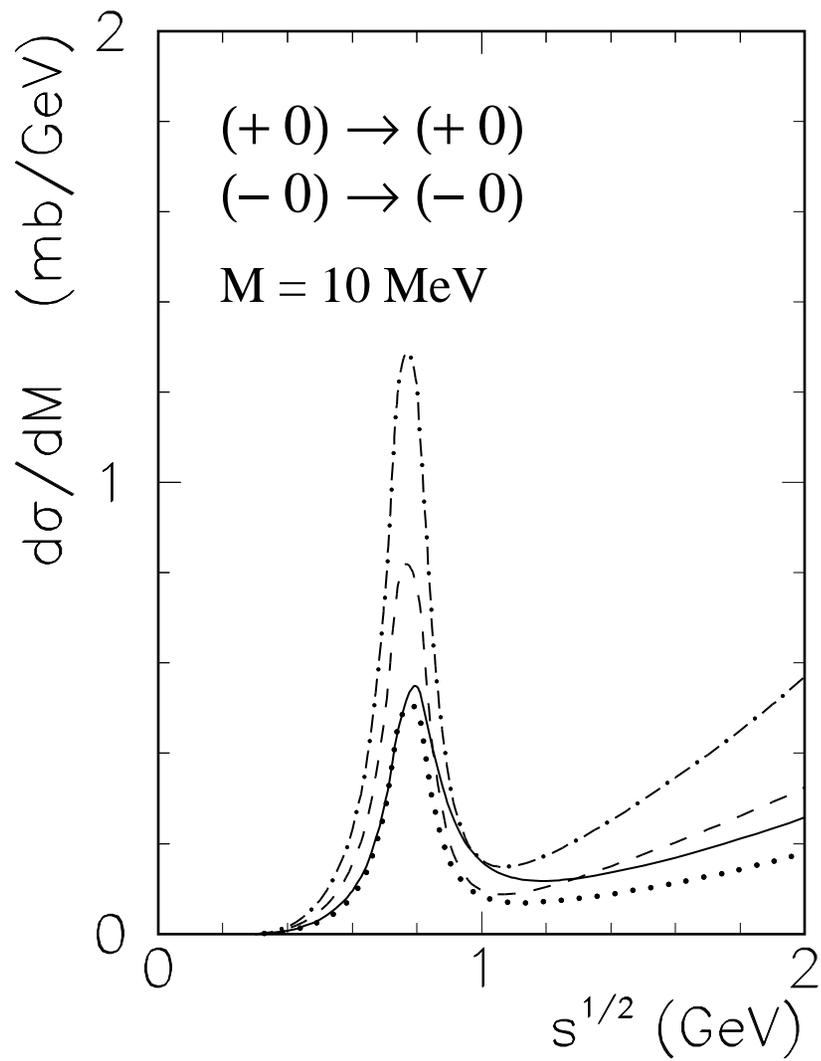
Figure 6:
Same as Figure 5, for the reactions
$\pi^+\pi^0 \to \pi^+\pi^0 e^+ e^-$ plus
$\pi^-\pi^0 \to \pi^-\pi^0 e^+ e^-$

\vspace{9mm}

\noindent
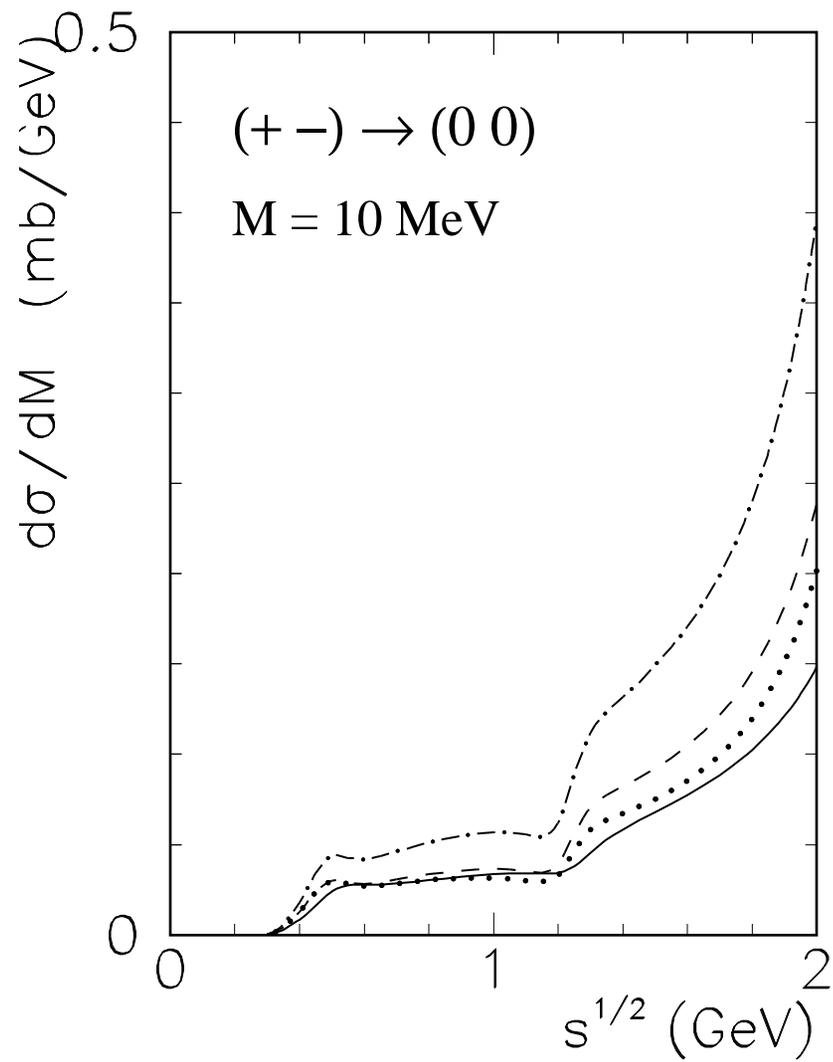
Figure 7:
Same as Figure 5, for the reaction
$\pi^+\pi^- \to \pi^0\pi^0 e^+ e^-$,
including final-state symmetry factor.
Note the different scales on the $y$-axes.

\vspace{9mm}

\noindent
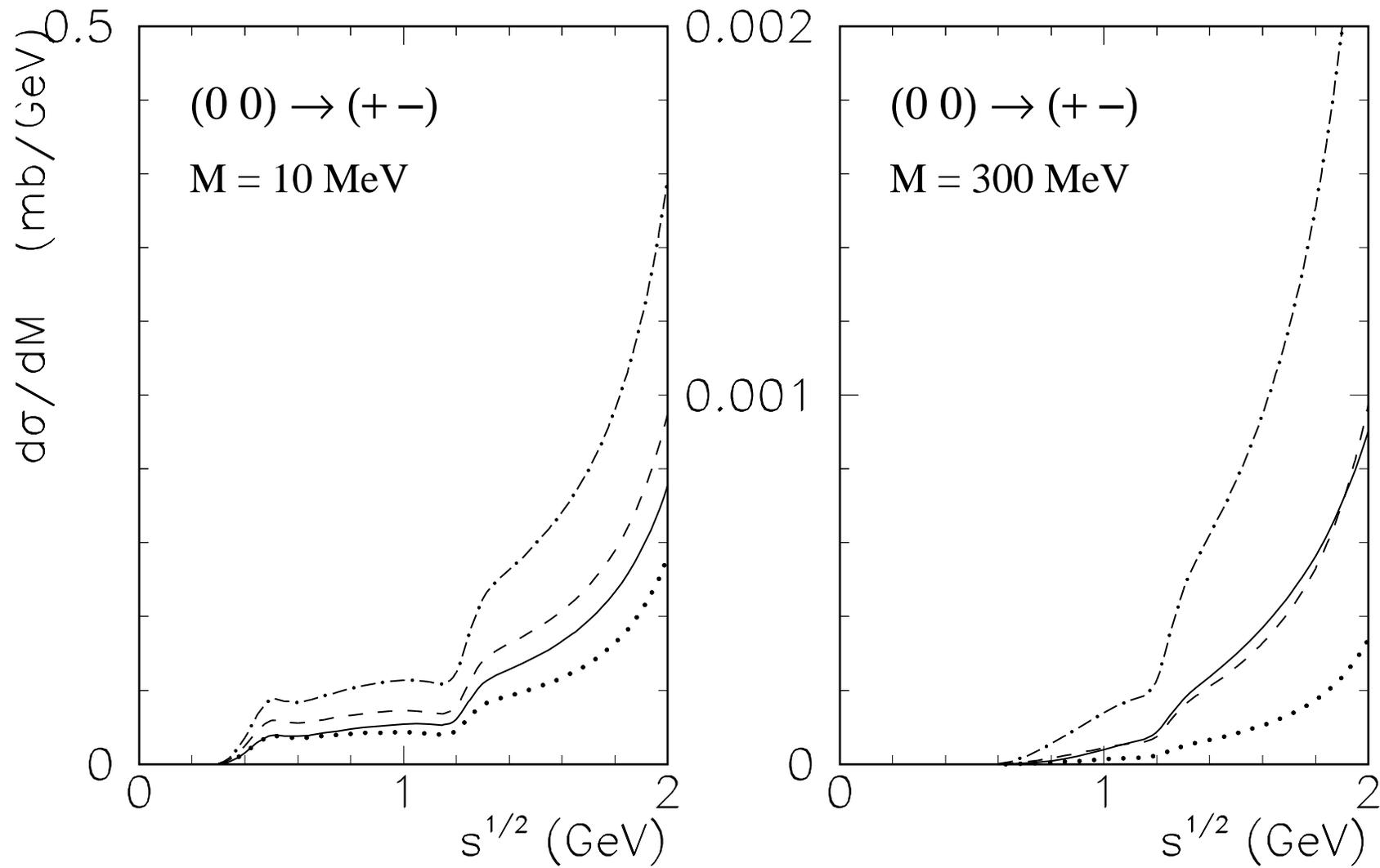
Figure 8:
Same as Figure 5, for the reaction
$\pi^0\pi^0 \to \pi^+\pi^- e^+ e^-$.
An initial-phase space factor of 1/2 is included.

\vspace{9mm}

\noindent
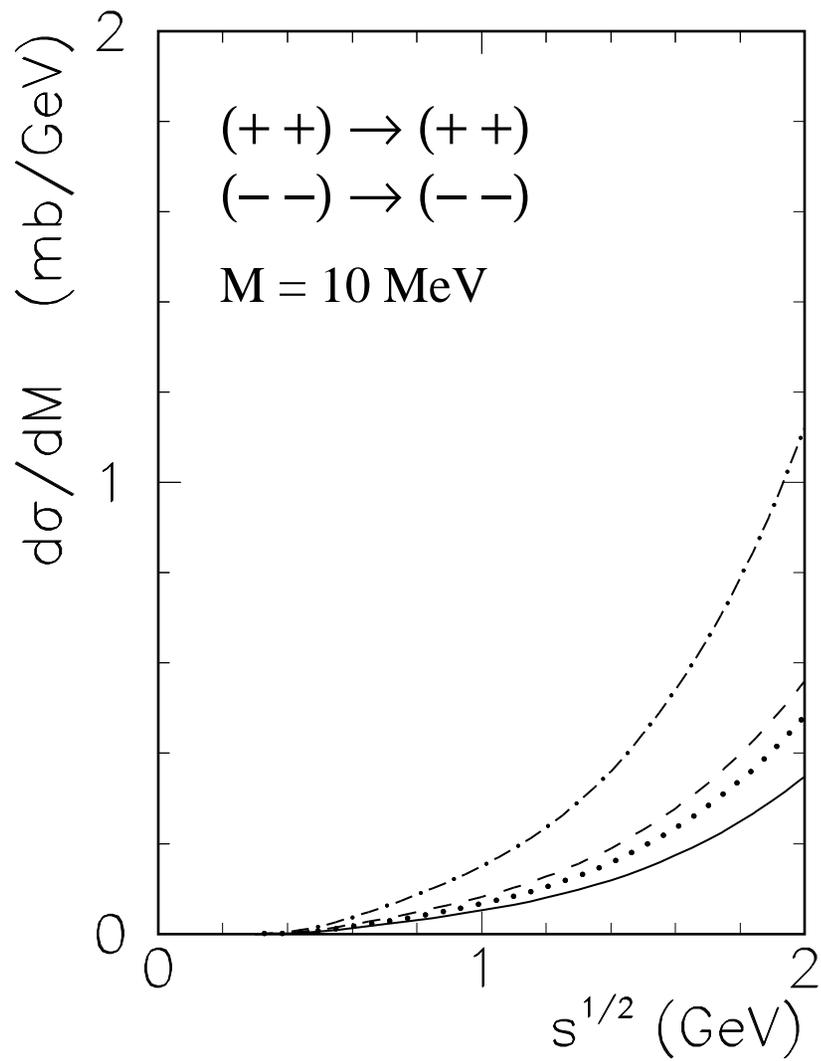
Figure 9:
Same as Figure 5, for the reactions
$\pi^+\pi^+ \to \pi^+\pi^+ e^+ e^-$ plus
$\pi^-\pi^- \to \pi^-\pi^- e^+ e^-$.
Initial- times final-state symmetry factors $(1/2)^2$ are included.

\vspace{9mm}

\noindent
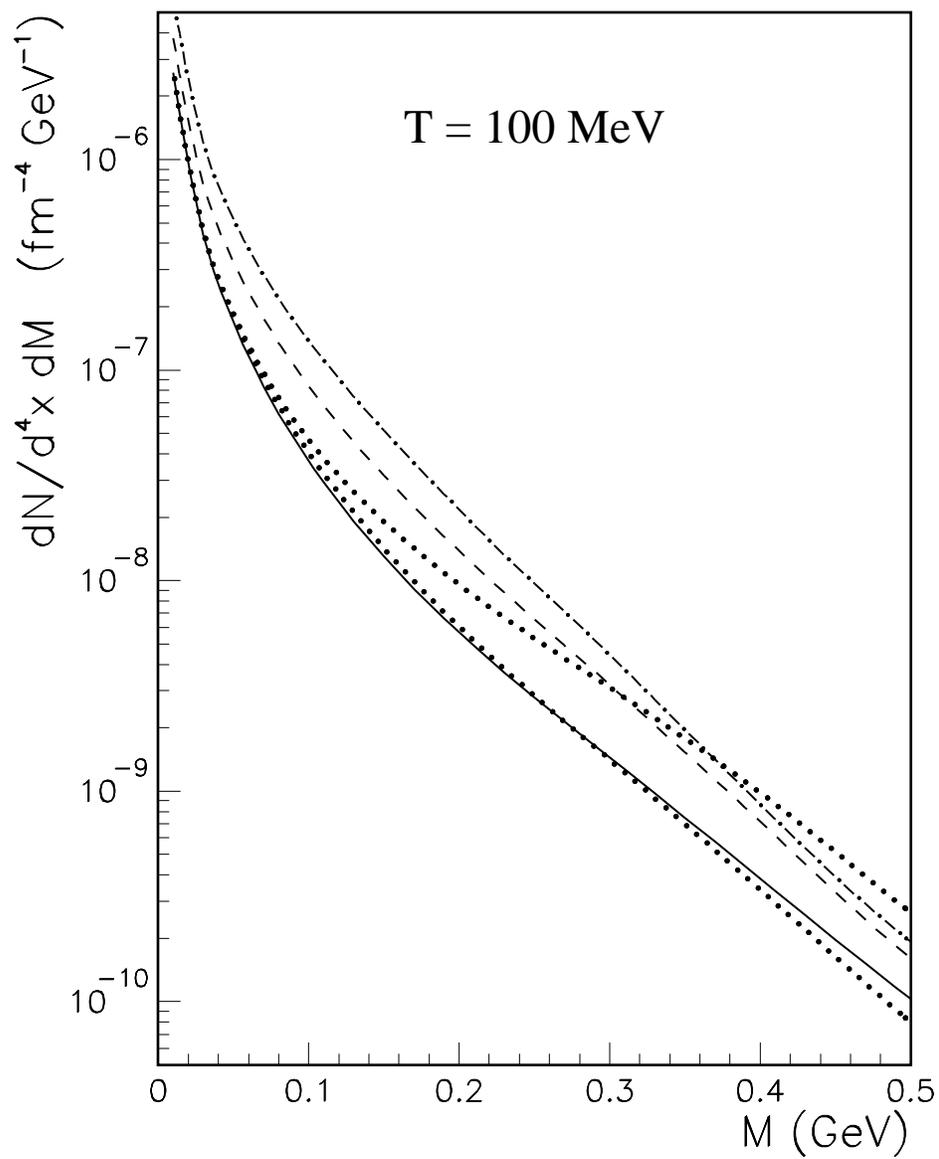
Figure 10:
Total bremsstrahlung yield of $e^+e^-$ for all seven
pion-pion reactions, for a Boltzmann gas with temperature
$T = 100$ MeV.
Solid line: exact OBE calculation,
Dash-dotted line: R\"uckl approximation,
Dashed line: angular-averaged virtual-$\gamma^*$ current,
Dotted line: 3-phase space current, virtual-$\gamma^*$ current (upper)
             real-$\gamma$ current (lower).

\vspace{9mm}

\noindent
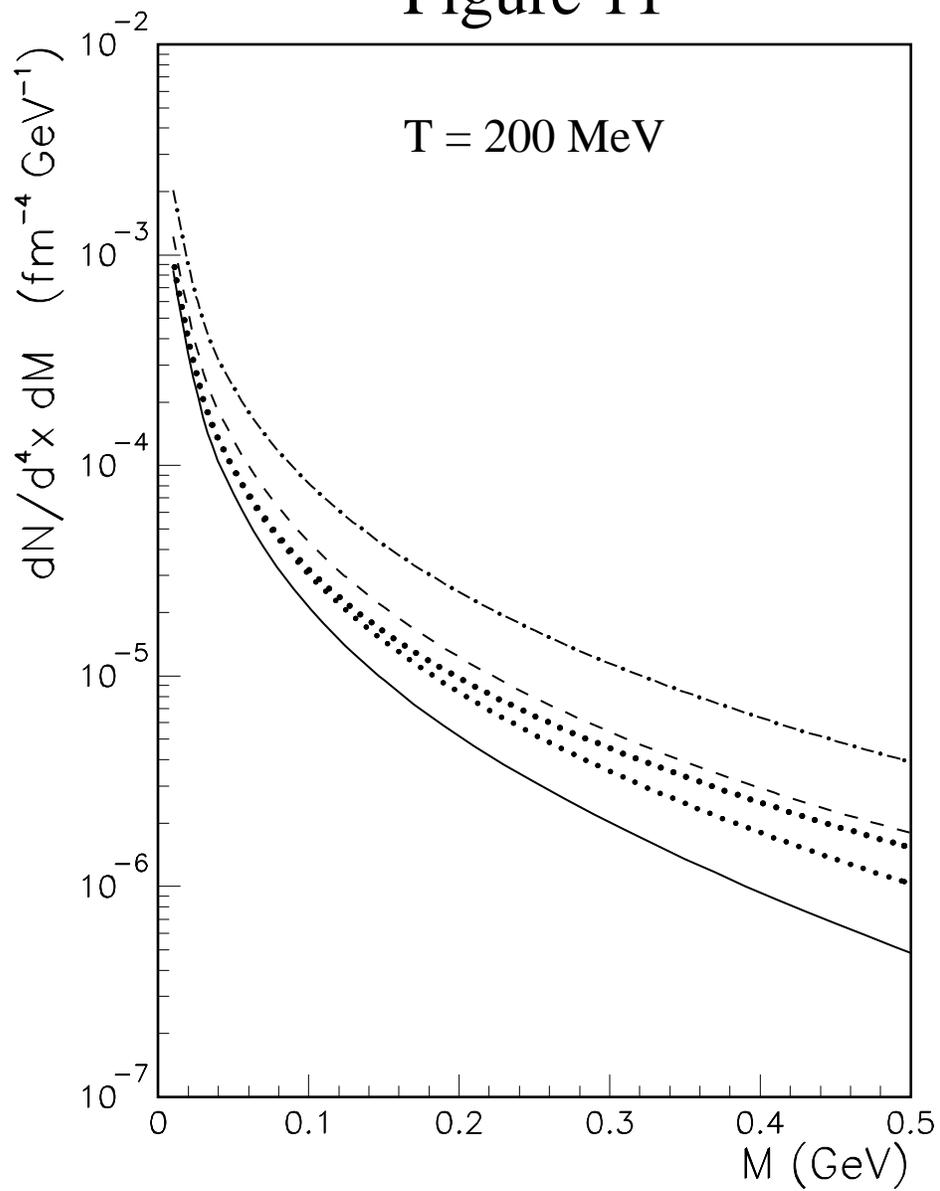
Figure 11:
Same as Figure 10, for temperature $T = 200$ MeV.

\vspace{9mm}

\noindent
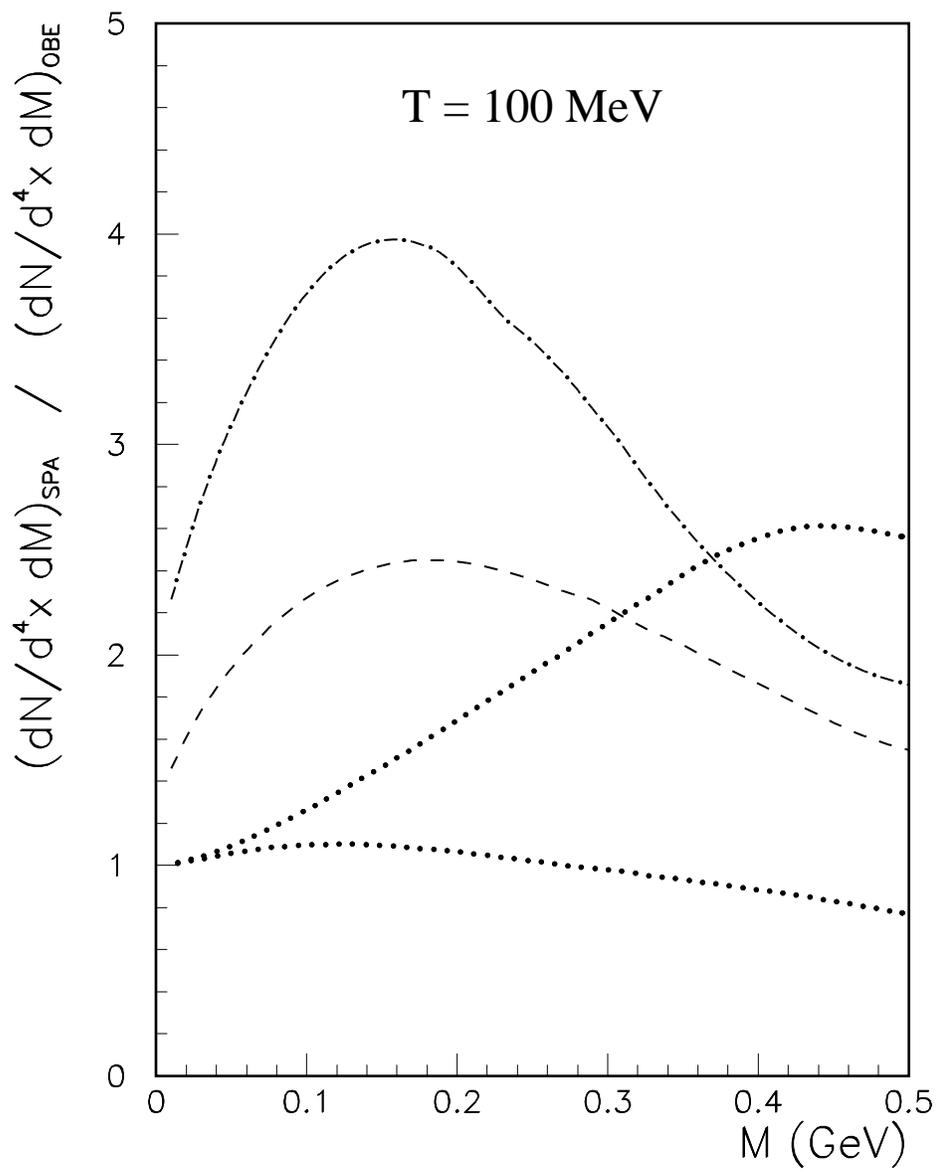
Figure 12:
Ratios of SPA approximation calculations divided by exact OBE rate
for $T = 100$ MeV.
Dash-dotted line: R\"uckl/exact,
Dashed line: (angular-averaged virtual $\gamma$ current)/exact.
Upper dotted line: (3-phase space virtual $\gamma$ current)/exact.
Lower dotted line: (3-phase space real $\gamma$ current)/exact.

\vspace{9mm}

\noindent
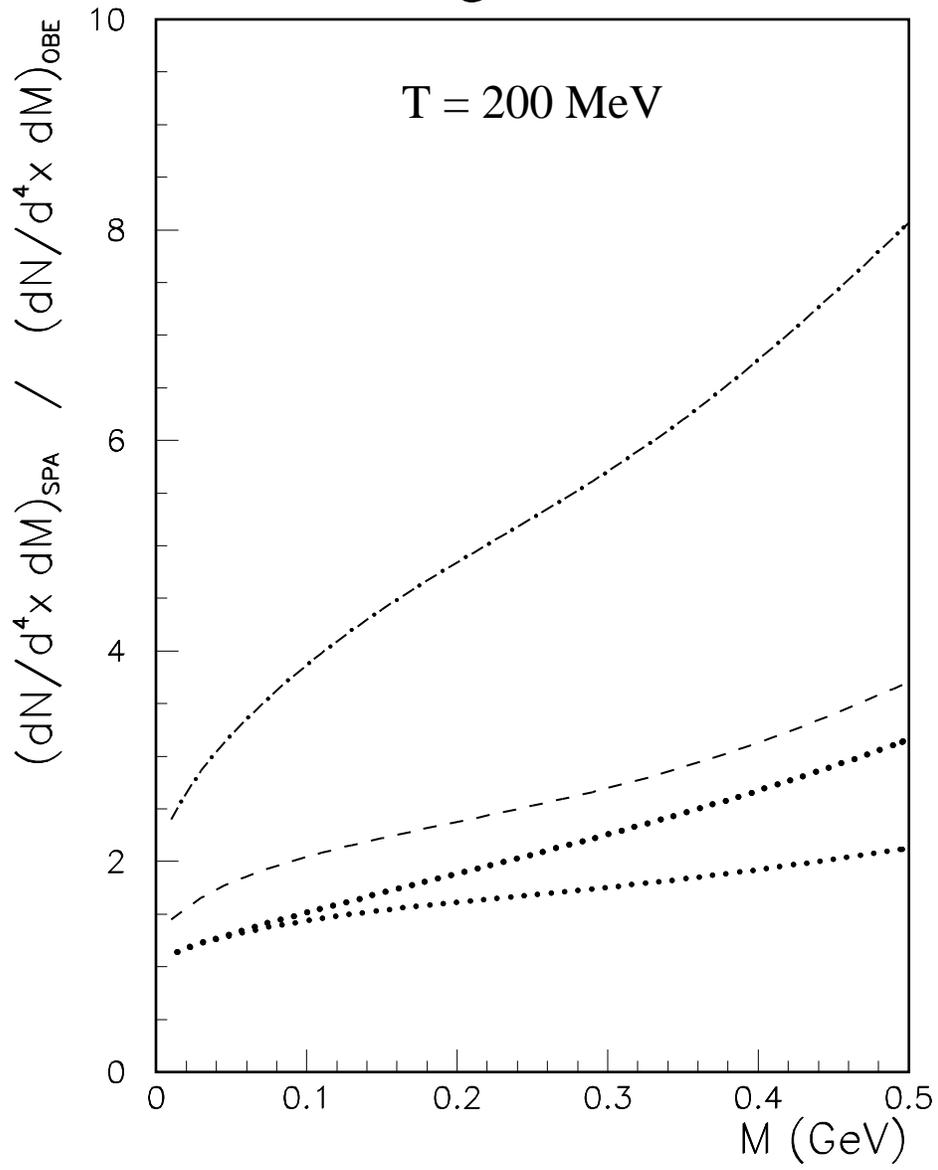
Figure 13:
Same as Figure 12, for $T = 200$ MeV,

\vspace{9mm}

\noindent
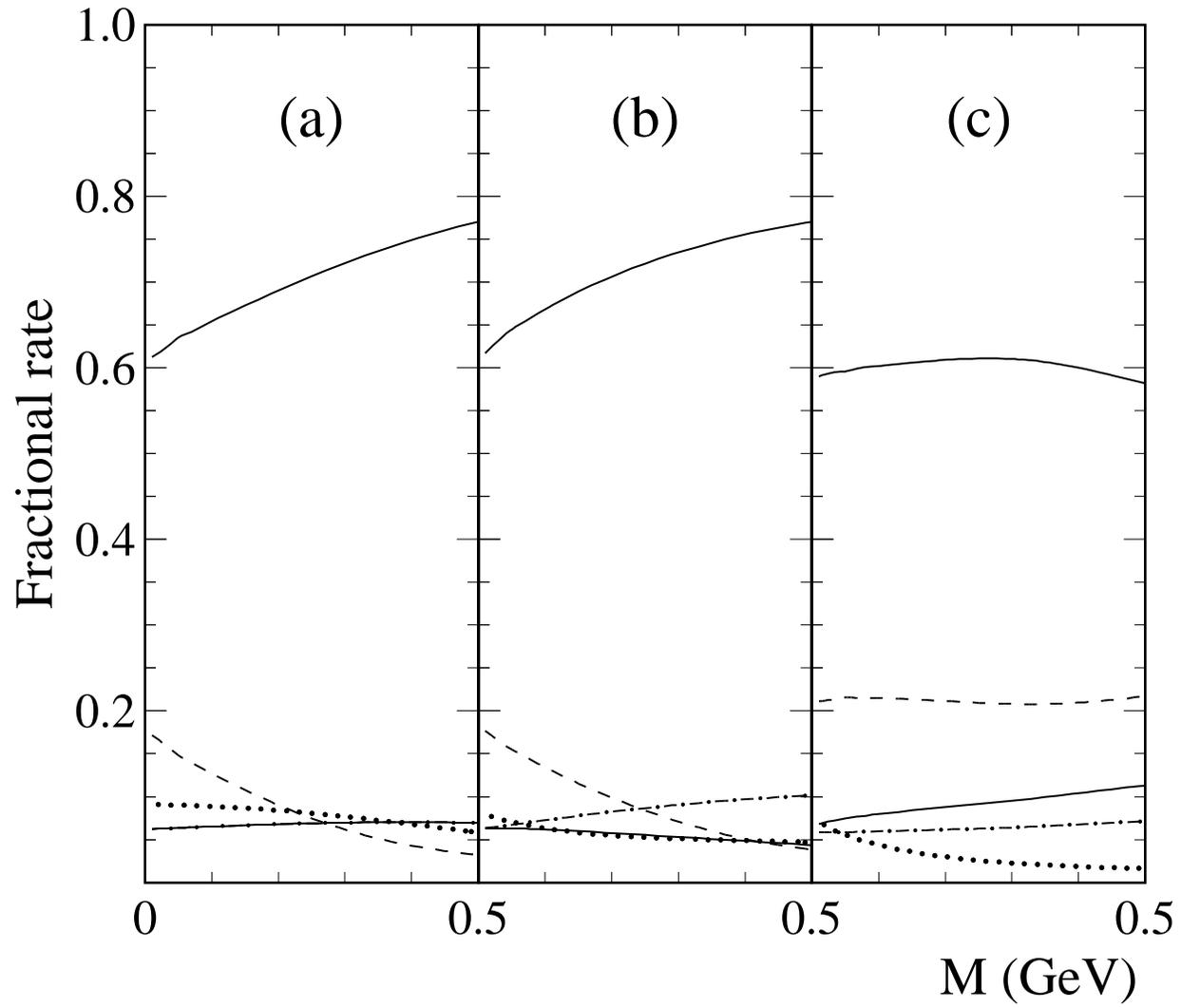
Figure 14:
Fractional contribution to the total dielectron rate at $T = 200$ MeV
as a function of dilepton invariant mass $M$
for (a) the R\"uckl, (b) virtual-photon Lichard approximations and
(c) the exact calculation.
Solid line (upper): $({+}{-}) \to ({+}{-})$,
Dashed line: $({+}{0}) \to ({+}{0})$ plus c.c.,
Solid line (lower): $({0}{0}) \to ({+}{-})$,
Dash-dotted line: $({+}{-}) \to ({0}{0})$,
Dotted line: $({+}{+}) \to ({+}{+})$ plus c.c.

\vspace{9mm}

\noindent
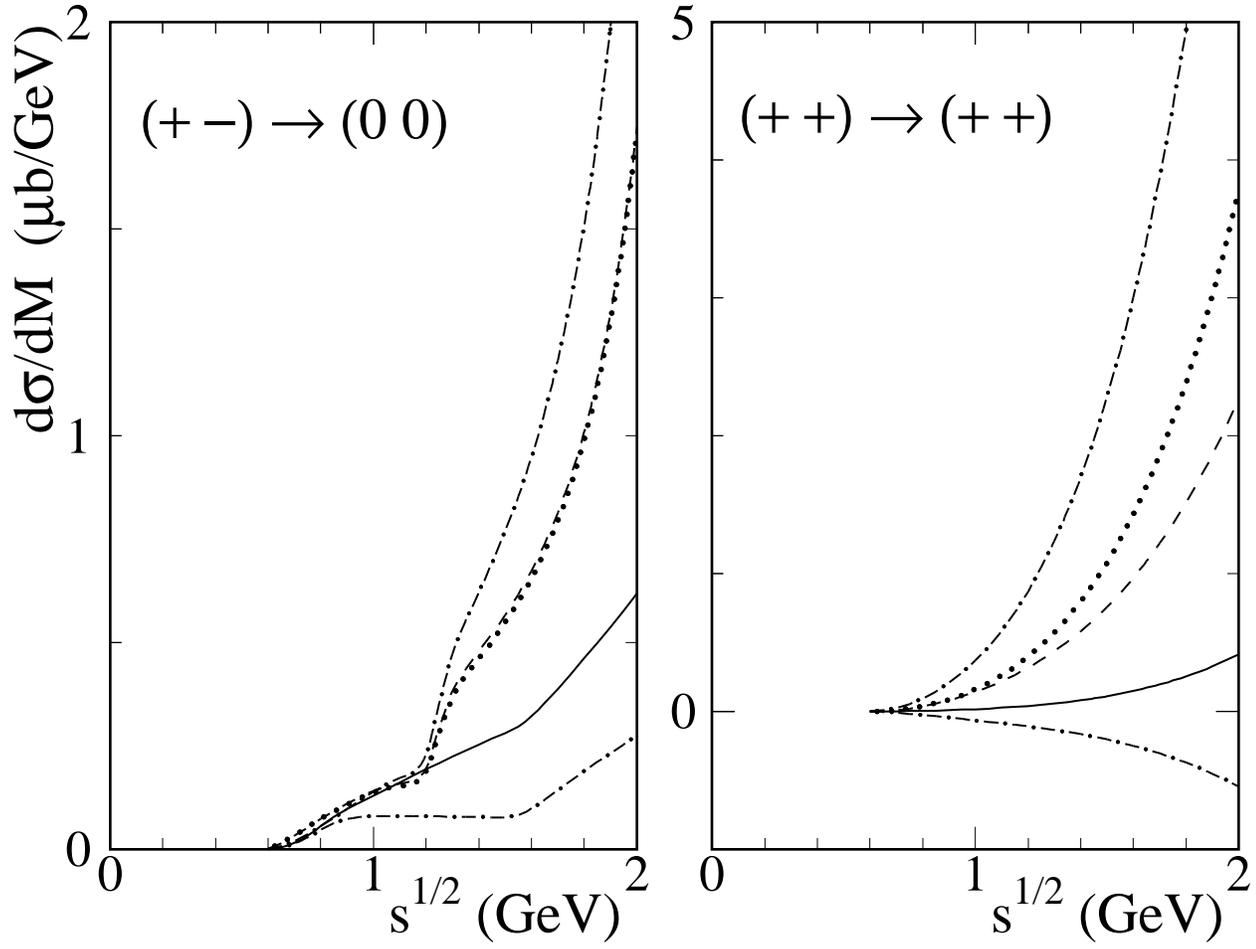
Figure 15:
Contribution of external-emission vs.\ external-plus-internal
diagrams for the reactions $({+}{-}) \to ({0}{0})$ (left)
and $({+}{+}) \to ({+}{+})$ (right), both for $M = 300$ MeV.
Lines are as in Figs.\ 7 and 9.
The new dash-dotted line below the (solid line) exact calculation
represents contributions arising solely from emission of
$\gamma^*$ by an external pion line, but taking $q$ into account
in ${\cal M}$, in contrast to the approximations (upper lines)
which neglected $q$.

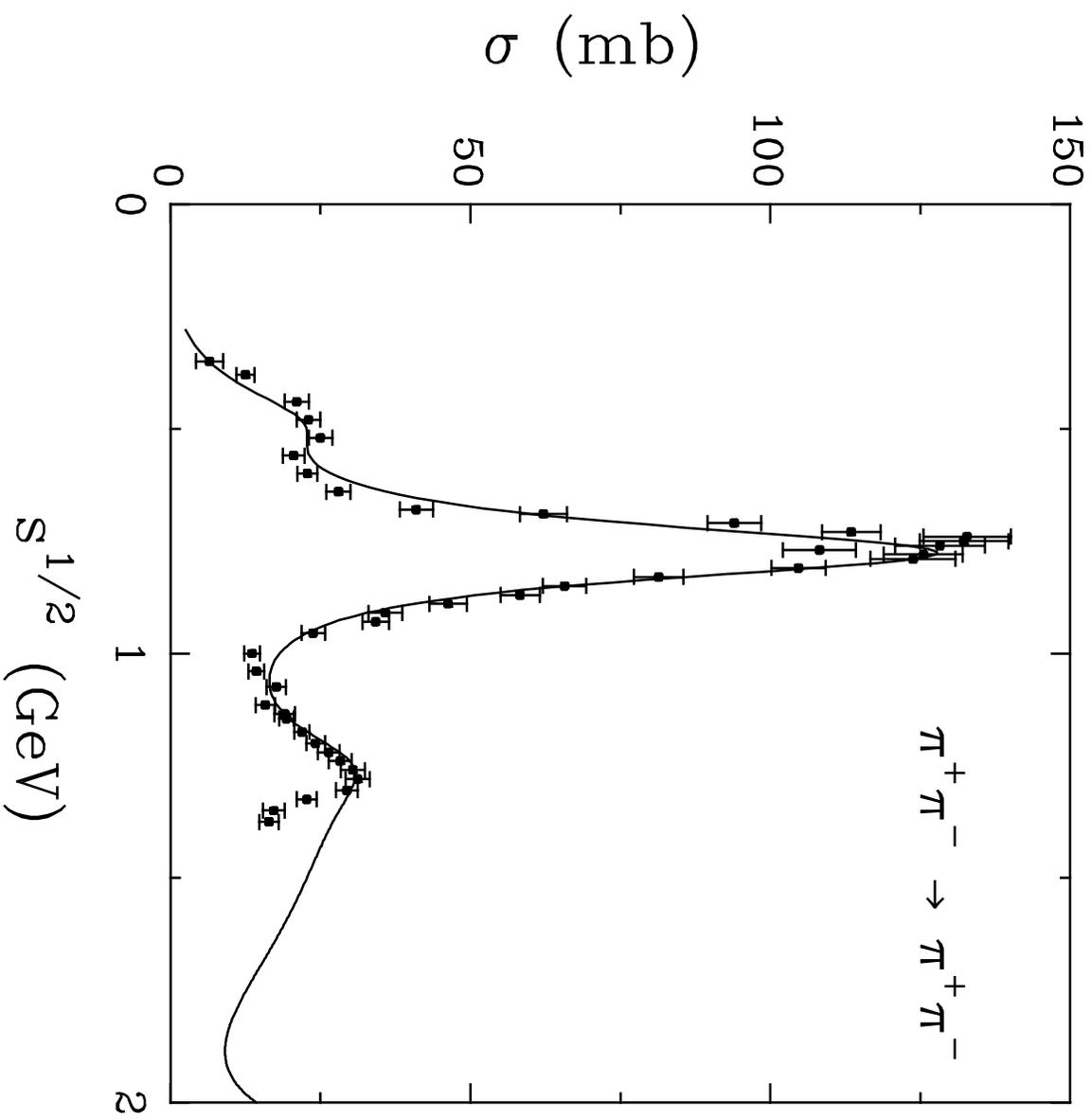
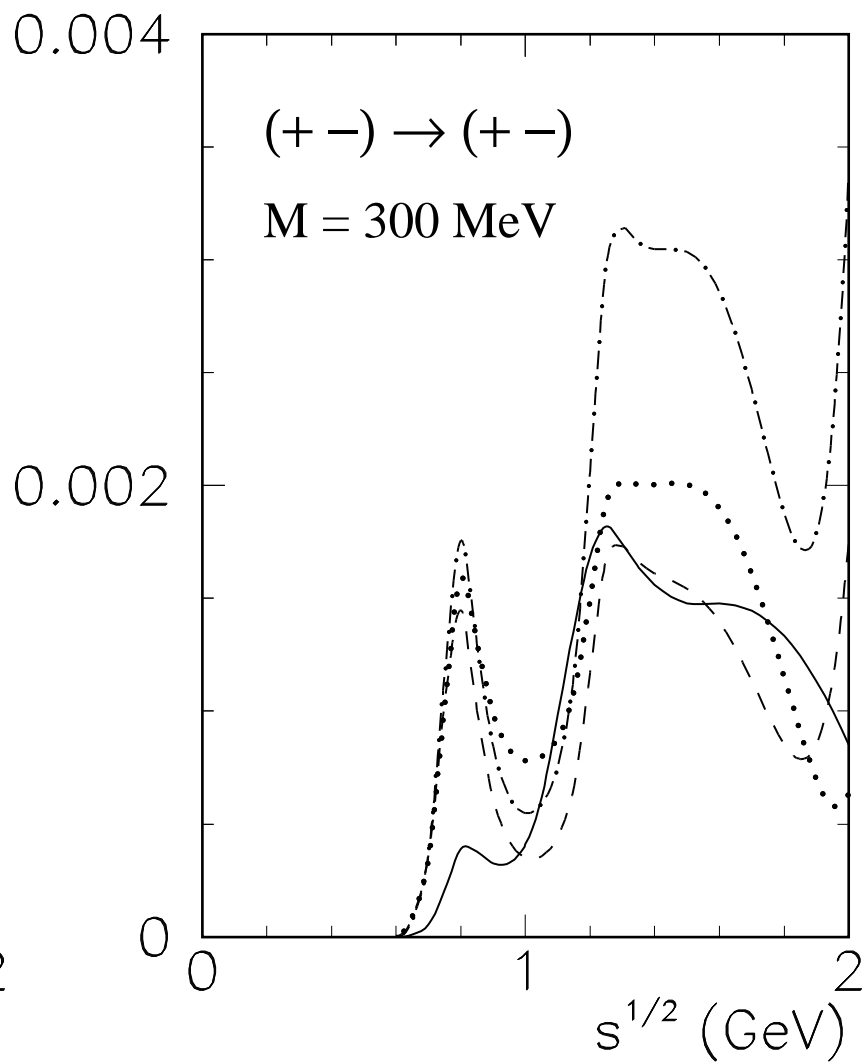
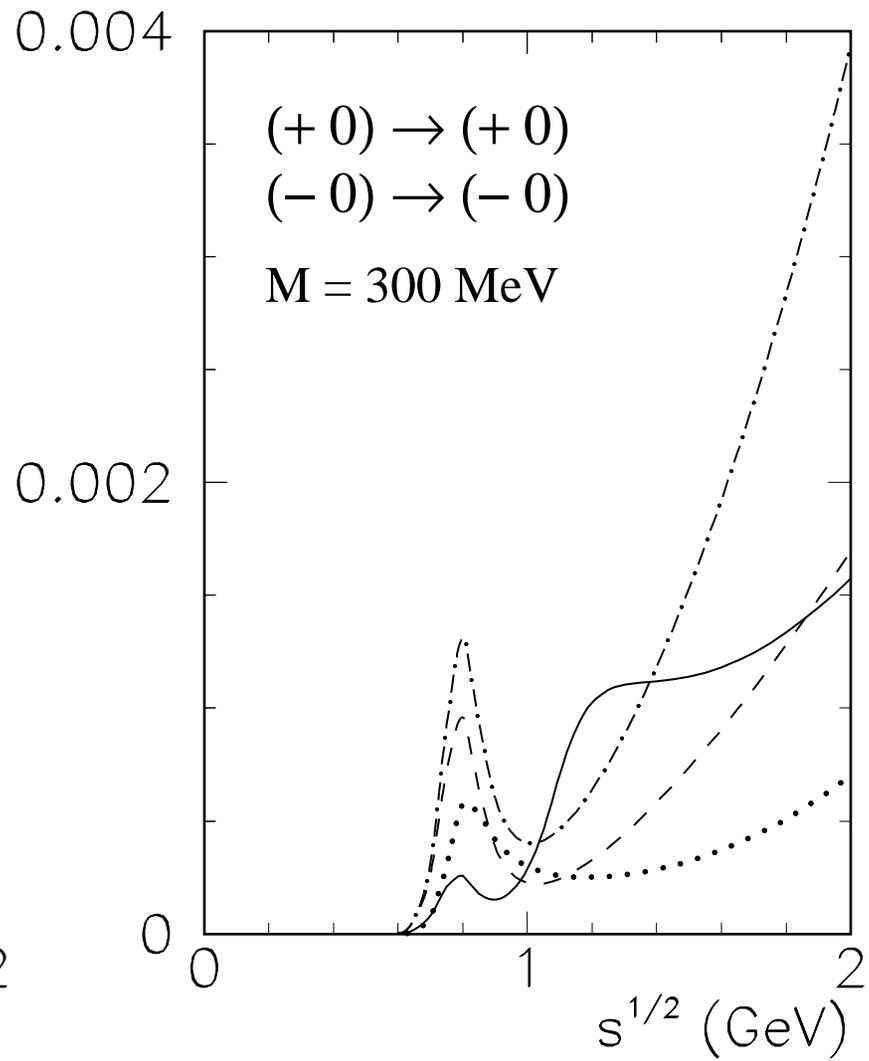
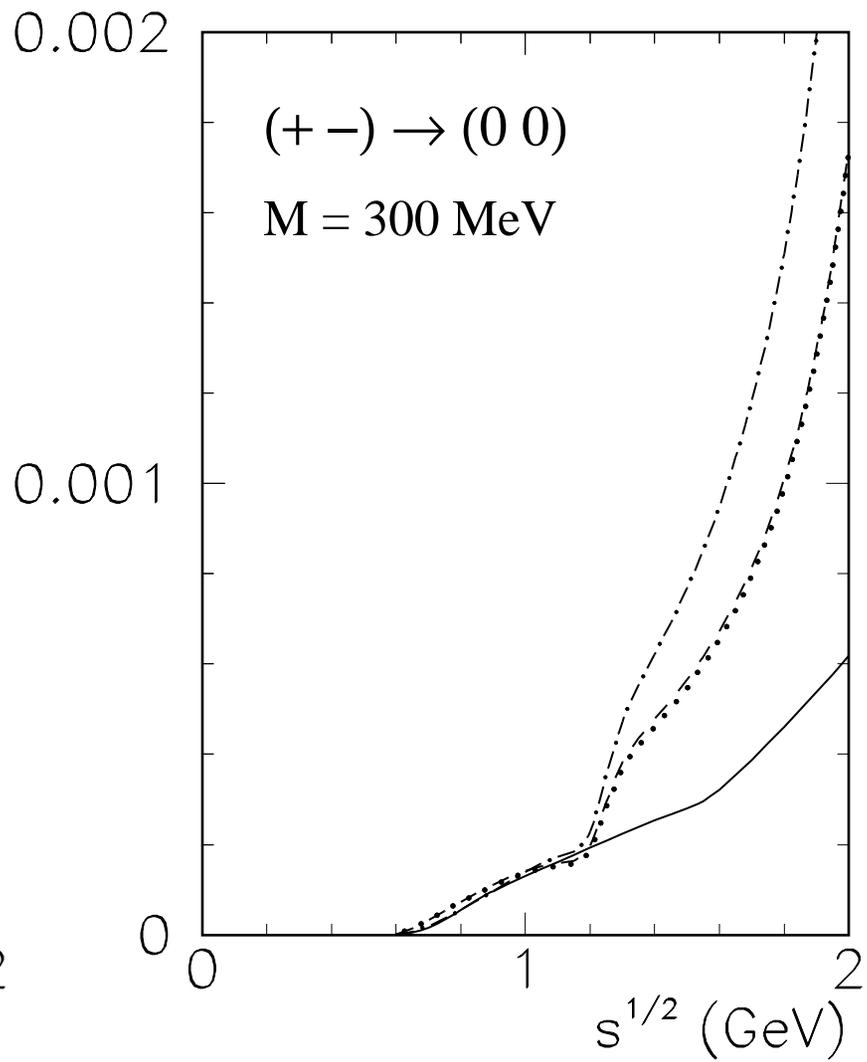
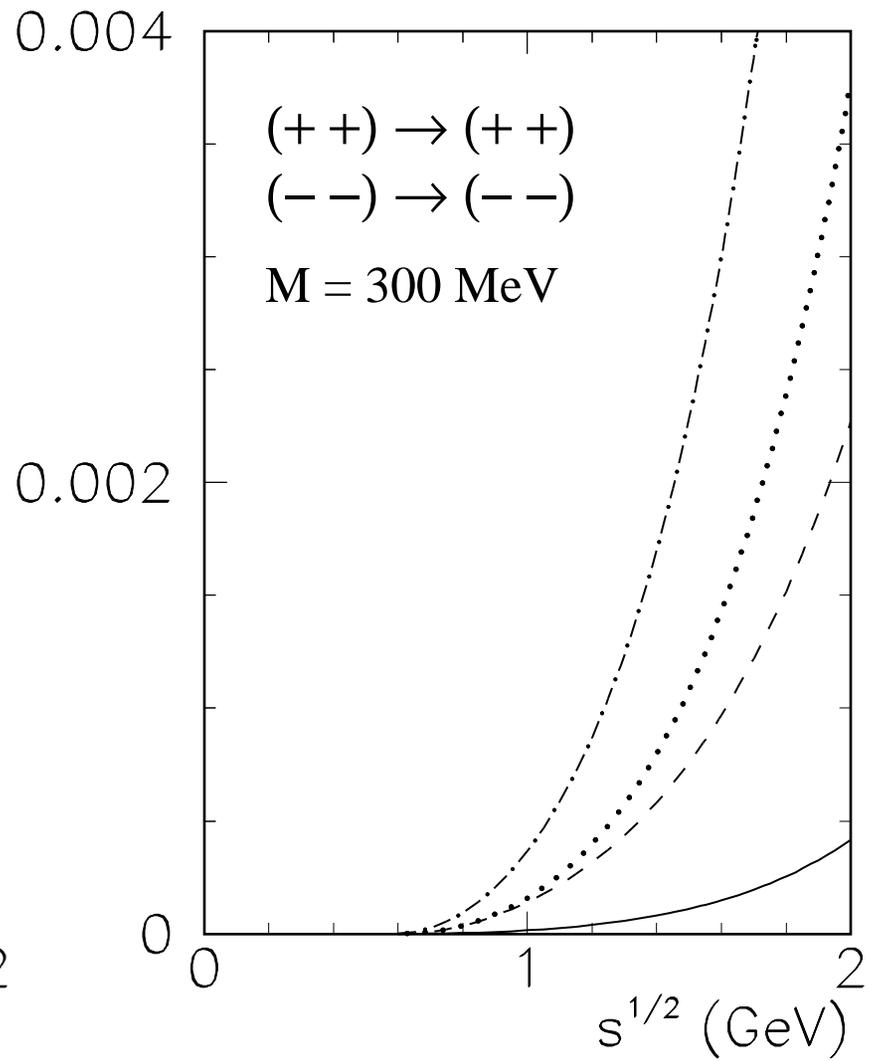
\end{document}